\documentclass[letterpaper]{ar-1col}
\usepackage{showyourwork}
\usepackage[letterpaper]{geometry}

\usepackage{natbib}
\usepackage{amsmath}
\usepackage{color}

\usepackage{hyperref}
\usepackage[version=4]{mhchem}
\usepackage{booktabs}

\newcommand{\mum}{$\mu$m}
\newcommand{\ld}{$\lambda/D$}
\newcommand{\fourier}[1]{\mathcal{F}\{#1\}}
\newcommand{\invfourier}[1]{\mathcal{F}^{-1}\{#1\}}
\newcommand{\acc}[1]{\entry{\acs{#1}}{\acl{#1}}}
\DeclareMathOperator*{\argmin}{arg\,min}
\usepackage{multicol}
\usepackage{etoolbox}

\usepackage{graphbox}

\usepackage{url}

\usepackage{totcount}
\usepackage[nolist,nohyperlinks]{acronym}

\begin{acronym}
    \acro{aplc}[APLC]   {Apodized Phase Lyot Coronagraph}
    \acro{paplc}[PAPLC] {Phase Apodized Phase Lyot Coronagraph}
    \acro  {psf}[PSF]   {Point Spread Function}
    \acro  {iwa}[IWA]   {Inner Working Angle}
    \acro  {owa}[OWA]   {Outer Working Angle}
    \acro  {ao}[AO]     {Adaptive Optics}
    \acro  {fpm}[FPM]   {Focal Plane Mask}
    \acro  {ppm}[PPM]   {Pupil Plane Mask}
    \acro  {hz}[HZ]     {Habitable Zone}
    \acro  {hlc}[HLC]   {Hybrid Lyot Coronagraph}
    \acro {fpwfs}[FPWFS]{Focal Plane Wavefront Sensing}
    \acro  {fwhm}[FWHM] {Full Width Half Maximum}
    \acro  {fqpm}[FQPM] {Four Quadrant Phase Mask}
    \acro  {agpm}[AGPM] {Annular Groove Phase Mask}
    \acro  {ovc}[OVC]   {Optical Vortex Coronagraph}
    \acro  {pp}[PP]     {Pupil Plane}
    \acro  {fp}[FP]     {Focal Plane}
    \acro  {sam}[SAM]     {Sparse Aperture Masking}
    \acro  {scar}[SCAR]     {Single-mode Complex Amplitude Refinement}
    \acro  {hci}[HCI]   {High Contrast Imaging}
    \acro  {dm}[DM]     {Deformable Mirror}
    \acro  {wfs}[WFS]   {Wavefront Sensor}
    \acro  {zwfs}[ZWFS] {Zernike Wavefront Sensor}
    \acro  {sphere}[SPHERE]{Spectro-Polarimetric High-contrast Exoplanet REsearch}
    \acro  {scfp}[SCFP] {Science Camera Focal Plane}
    \acro  {smd}[SMD]   {Spatial Mode Demultiplexing}
    \acro  {ncpa}[NCPA] {Non-Common Path Aberration}
    \acro  {mspl}[MSPL] {Mode-selective Photonic Lantern}
    \acro  {elt}[ELT]   {Extremely Large Telescope}
    \acro  {jwst}[JWST] {James Webb Space Telescope}
    \acro  {gmt}[GMT]   {Giant Magellan Telescope}
    \acro  {tmt}[TMT]   {Thirty Metre Telescope}
    \acro  {spp}[SPP]     {Shaped Pupil Plate}
    \acro  {efc}[EFC]     {Electric Field Conjugation}
    \acro  {scc}[SCC]   {Self Coherent Camera}
    \acro  {piaa}[PIAA] {Phase Induced Amplitude Apodization}
    \acro  {piaacmc}[PIAACMC] {Phase Induced Amplitude Apodization Complex Mask Coronagraph}
    \acro  {fast}[FAST] {Fast Atmospheric SCC}
    \acro  {smscc}[SM-SCC] {Spectrally Modulated SCC}
    \acro  {app}[APP]   {Apodizing Phase Plate}
    \acro  {gvapp}[gvAPP]   {grating vector Apodizing Phase Plate}
    \acro  {vvc}[VVC]   {Vector Vortex Coronagraph}
   \acro  {vfn}[VFN]    {Vortex Fibre Nulling}
   \acro  {ravc}[RAVC]  {Ring Apodized Vortex Coronagraph}
    \acro  {hwo}[HWO]   {Habitable Worlds Observatory}
    \acro  {hci}[HCI]   {High Contrast Imaging}
    \acrodefplural{psf}[PSFs]   {Point Spread Functions}
    \acrodefplural{dm}[DMs]     {Deformable Mirrors}
    \acrodefplural{wfs}[WFSs]   {Wavefront Sensors}
    \acrodefplural{elt}[ELTs]   {Extremely Large Telescopes}
    \acrodefplural{ncpa}[NCPAs] {Non-Common Path Aberrations}
\end{acronym}

\newtotcounter{citnum} 
\def\oldbibitem{} \let\oldbibitem=\bibitem
\def\bibitem{\stepcounter{citnum}\oldbibitem}

\jname{Annu. Rev. Astron. Astrophys.}
\jvol{63}
\jyear{2025}
\doi{10.1146/annurev-astro-021225-022840}

\newcommand{\project}[1]{\textsf{#1}}
\hypersetup{%
        colorlinks,
        breaklinks=true,
        plainpages=false,%
        citecolor=[rgb]{0,0.20,0.45},
        linkcolor=[rgb]{0,0.20,0.45},
        urlcolor=[rgb]{0,0.20,0.45},
        bookmarksopen=true,%
        bookmarksnumbered=false,%
        bookmarksdepth=5%
}

\begin{document}

\markboth{Kenworthy \& Haffert}{HCC}

\title{\huge {\normalfont High-Contrast Coronagraphy}}

\author{\large {\normalfont Matthew A. Kenworthy$^1$ and Sebastiaan Y. Haffert$^{1,2}$}
  \affil{$^1$Leiden Observatory, Niels Bohrweg 2, Leiden 2300RA, The Netherlands; email: kenworthy@strw.leidenuniv.nl}
  \affil{$^2$University of Arizona, Steward Observatory, 933 N. Cherry Avenue, Tucson, Arizona, United States; email: haffert@strw.leidenuniv.nl}}

\begin{abstract}
Imaging terrestrial exoplanets around nearby stars is a formidable technical challenge, requiring the development of coronagraphs to suppress the stellar halo of diffracted light at the location of the planet.
In this review, we derive the science requirement for high-contrast imaging, present an overview of diffraction theory and the Lyot coronagraph, and define the parameters used in our optimization.
We detail the working principles of coronagraphs both in the laboratory and on-sky with current high-contrast instruments, and we describe the required algorithms and processes necessary for terrestrial planet imaging with the extremely large telescopes and proposed space telescope missions:

\begin{itemize}
    \item Imaging terrestrial planets around nearby stars is possible with \\ 
    a combination of coronagraphs and active wavefront control \\
    using feedback from wavefront sensors.
    \item Ground based 8-40m class telescopes can target the habitable \\ 
    zone around nearby M dwarf stars with contrasts of $10^{-7}$ and \\
    space telescopes can search around solar-type stars with \\
    contrasts of $10^{-10}$.
    \item Focal plane wavefront sensing, hybrid coronagraph designs and \\
    multiple closed loops providing active correction are required \\
    to reach the highest sensitivities.
    \item Polarization effects need to be mitigated for reaching $10^{-10}$ \\ contrasts whilst keeping exoplanet yields as high as possible.
    \item Recent technological developments, including photonics and micro-\\
    wave kinetic inductance detectors,
    will be folded into high-contrast \\ instruments.
\end{itemize}

\end{abstract}

\begin{keywords}
optics, coronagraphs, exoplanets, high contrast, computational methods
\end{keywords}
\maketitle

\tableofcontents

\section{Introduction}
\label{sec:intro}

Initially developed to image the Sun's corona without the need for a Solar eclipse \citep{Lyot33}, one of the most significant science drivers for the latest coronagraphs involves the detection and characterisation of circumstellar material and planets around nearby stars.
Young self-luminous gas giant exoplanets have been directly imaged at infrared wavelengths around young stars \citep[see ][ for a review of these detections]{Zurlo24} both in the nearby Galactic field and further away in young stellar OB associations (out to 400 pc), typically these exoplanets have luminosities of $10^{-4}-10^{-6}$ of their parent star at angular separations up to a few arcseconds. 

The search for life beyond the Earth has focused on the idea that water is an essential part of life cycles elsewhere in the Universe, as it is a polar solvent formed from two elements that are found in abundance throughout the Galaxy.
Places where water can exist in its liquid state form prime locations for these searches, notably Earth-like planets and ice moons that have a liquid water ocean underneath an ice layer.
The region around a star where liquid water can exist on the surface of a terrestrial planet (with appropriate atmospheric pressure) is referred to as the \ac{hz}; for the Sun this is from 0.9-1.2 au but this can move as the luminosity of stars evolve over time.

\begin{figure}[ht]
  \centering
  \script{plot_flux_ratio.py}
  \includegraphics[width=1.0\linewidth]{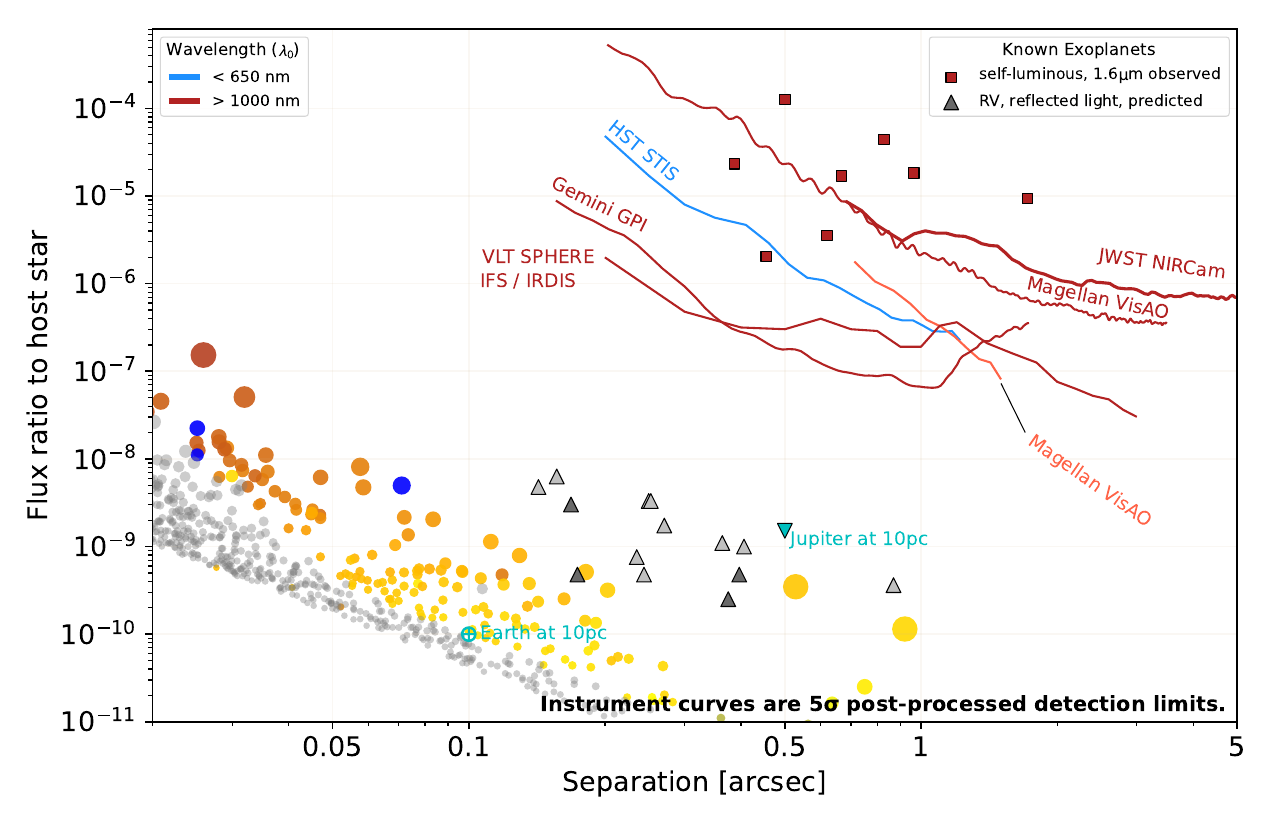}
  \caption{Angular separation versus contrast for directly imaged exoplanets.
  Known self-luminous gas giant planets are shown in the top right \citep{Lacy20}.
  The lines represent $5\sigma$ point source contrast limits from recent instruments and surveys.
  Figure plot and data from \citet{Bailey24} along with references for the contrast curves.
  Grey triangles are the expected contrasts for reflected light from known radial velocity detected exoplanets \citep{Bataltha18}.
  Circles represent reflected light estimates for 1 Earth radius planets, assuming one planet per star within 20 parsecs of the Sun in the Habitable Zone at maximum elongation.
  Grey points are planets that are fainter than 30th magnitude in $V$.
  The color represents the effective temperature of the star.
  The larger the point, the closer the stellar system is to the Sun.
  Reflected light data from \href{https://subarutelescope.org/staff/guyon/04research.web/14hzplanetsELTs.web/catalog.web/content.html}{Guyon (2024) web page}.  
}
  \label{fig:nstars}
\end{figure}

At visible light wavelengths, the flux from exoplanets that can be directly imaged is overwhelmingly dominated by the reflected light from their parent star.
For an Earth analogue with similar radius, albedo and effective temperature orbiting around a solar-type star 10 parsecs away, the typical amount of reflected light in the optical is $10^{-10}$ of the central star at a separation of 0.1 arcseconds at maximum elongation.
The technical challenge is in distinguishing the light of the parent star from the light of the planet.
Planetary systems that are closer to the Sun have two benefits: one that is twice as close as the other will have double the angular separation between the star and planet, and from the inverse square law, four times more flux is received from the planet.
For direct imaging, therefore, the closest stars to the Sun are the ones that are studied for the presence of directly imaged exoplanets.
On average there are more M dwarf stars close to the Sun than other solar type stars: in a volume limited (20 parsec) sample around the Sun, there are $\sim 140$ solar G-type stars, and on the order of 1900 M-dwarf stars \citep{Kirkpatrick24}.
For solar type stars, the contrast required in the optical wavelengths is on the order of $10^{-10}$ for terrestrial planets in the \ac{hz}, but for smaller mass stars with lower luminosities, the contrast is $10^{-7}$ for M dwarfs - for stars around the Sun this is shown in Figure~\ref{fig:nstars}.

The markers for biosignatures also set the parameters (such as wavelength and bandwidth) that form part of the design decision.
Spectra of Earthshine (sunlight that is reflected off the Earth and illuminates the dark portion of the Moon) from 0.5 to 2.4 microns show \ce{O2}, \ce{O3}, \ce{CO2}, \ce{CH4} and \ce{H2O} \citep{Turnbull06}, and there are many discussions about the possible biomarker molecules that should be searched for and characterised as evidence for biosignatures \citep[see the reviews of ][]{2016AsBio..16..465S,2017ARAA..55..433K,2018AsBio..18..663S}.
It is clear that the unambiguous detection of life will require not one singular detection of a molecule, but will be the combination of several different lines of evidence.
We focus on the technical challenges for imaging a rocky terrestrial planet around a nearby star in our Galaxy.

\section{From Maxwell's Equations to Wavefronts}\label{sec:maxwell}

\begin{table}
\caption{Definition of physical quantities}\label{tab1}
\begin{center}
\begin{tabular}{@{}l|l|l|l@{}}
\hline
$i$ & The imaginary unit & $c$ & Speed of light in a vacuum \\\hline
$\mathcal{E}$ & Electric field & $\mathcal{H}$ & Magnetic field \\\hline
$\mathcal{D}$ & Electric displacement & $\mathcal{B}$ & Magnetic induction \\\hline
$\epsilon$ & Relative electric permittivity & $\epsilon_0$ & Electric permittivity of vacuum \\\hline
$\mu$ & Relative magnetic permeability & $\mu_0$ & Magnetic permeability of vacuum \\\hline
$\omega$ & Angular frequency & $|\vec{k}|$ & Wave number \\\hline
$\lambda$ & Wavelength & $\lambda_0$ & Central wavelength in the bandpass \\\hline
$\Pi$ & The telescope pupil function & $\Delta\lambda$ & The width of the bandpass \\\hline
$D$ & Diameter of the pupil & $n$ & (Complex) refractive index of macroscopic media \\\hline
$\mathcal{F}_{x,y}[\cdot]$ & Fourier transform operator & $\mathcal{F}^{-1}_{x,y}[\cdot]$ & Inverse Fourier transform operator \\\hline
$C_\lambda[\cdot]$ & A general coronagraph propagation operator & $\Psi_\lambda[\vec{k}]$ & Coronagraphic image \\\hline
$\phi$ & The phase of the electric field & $\alpha$ & The amplitude of the electric field \\\hline
\hline
\end{tabular}
\end{center}
\end{table}

The vast majority of energy from astrophysical objects arrives at our telescopes in the form of electromagnetic radiation.
This time-dependent interaction is described by Maxwell's equations \citep[with relevant physical quantities defined in Table \ref{tab1} and following ][]{Lavrinenko14},

\begin{equation}
\begin{aligned}
\frac{\partial\mathcal{D}}{\partial t} \quad & = \quad \nabla\times\mathcal{H},   & \quad \text{(Faraday's law)} \\[5pt]
\frac{\partial\mathcal{B}}{\partial t} \quad & = \quad -\nabla\times\mathcal{E},  & \quad \text{(Ampere's Law)}   \\[5pt]
\nabla\cdot\mathcal{B}                 \quad & = \quad 0,                         & \quad \text{(Gauss's law)}   \\[5pt]
\nabla\cdot\mathcal{D}                 \quad & = \quad 0.                         & \quad \text{(Coulomb's law)}
\end{aligned}
\end{equation}

This form of Maxwell's equations is in the material form without any charge and current sources.
The material form is used to describe the propagation of electromagnetic fields inside matter.
This set of equations is completed by describing the particular matter of the medium with the constitutive relations, $\mathcal{D}=\epsilon \mathcal{E}$ and $\mathcal{B}=\mu \mathcal{H}$. 
Here $\epsilon$ and $\mu$ are the permittivity and magnetic permeability, respectively.
The wave equation for electromagnetic waves can be derived by taking the curl of Ampere's law.
This gives us the classic wave equation if we assume that the EM wave propagates in isotropic and homogeneous materials,
\begin{equation}
\label{eq:wave_eq}
\begin{aligned}
\mu \epsilon\frac{\partial^2 \mathcal{E}}{\partial t^2} - \nabla^2\mathcal{E} = 0 .
\end{aligned}
\end{equation}
We consider a purely monochromatic EM wave with angular frequency $\omega$.
The wave function is then $\mathcal{E}=\psi(r) e^{iwt}$ with $\psi(r)$ describing the spatial distribution.
Substituting this relation into Equation \ref{eq:wave_eq},
\begin{equation}
\nabla^2\mathcal{E} +\mu \epsilon \omega^2 \mathcal{E} = 0.
\end{equation}
The usual definition of the permittivity and permeability are $\epsilon=\epsilon_r \epsilon_0$ and $\mu=\mu_r \mu_0$ with $\epsilon_r$ and $\mu_r$ the relative permittivity and permeability compared to that of vacuum, $\epsilon_0$ and $\mu_0$.
In optics most glasses and materials are defined by their refractive index $n$ which is related to permittivity as $\epsilon_r = n^2$ and many of these are also non-magnetic which means that $\mu_r=1$.
Substituting this will give us the classic Helmholtz equation
\begin{equation}
\nabla^2\mathcal{E} + n^2k^2 \mathcal{E} = 0,
\end{equation}
We made use of the fact that the speed of light is $c = \frac{1}{\sqrt{\epsilon_0\mu_0}}$ and that $k = c\omega$ is the wave number.
The propagation through an optical system has a preferential direction that is usually defined along the z-axis.
The z-axis evolution can be derived by separating the spatial components into the z component and the perpendicular components (x,y),
\begin{equation}
\frac{\partial^2\mathcal{E}}{\partial z^2} = -\nabla_{\perp}^2\mathcal{E}-n^2k^2 \mathcal{E},
\end{equation}
This differential equation can be solved by assuming a plane wave expansion $\mathcal{E}(x,y,z)=e^{i(k_x x + k_y y)}f(z)$ which results in
\begin{equation}
\frac{\partial^2\mathcal{E}}{\partial z^2} = -(n^2k^2 - k_{\perp}^2)\mathcal{E}.
\end{equation}
The solution to this equation is the so called Angular Spectrum Propagator that relates the electric field at any one plane to the electric field at any other,
\begin{equation}
\label{eq:angular_spectrum}
\mathcal{E}(x, y, z') = \mathcal{F}_{x,y}^{-1}\{e^{-ik_z(z'-z)}\mathcal{F}_{x,y}\{\mathcal{E}(x,y,z)\}\}.
\end{equation}

Here the $z$ component of the wave vector is defined as $k_z=\sqrt{n^2k^2 - k_{\perp}^2}$ and $\mathcal{F}_{x,y}^{(-1)}$ is defined as the (inverse) Fourier transform over the x and y coordinates.
While Equation \ref{eq:angular_spectrum} describes the full propagation from one plane to another, it is quite unwieldy to use and does not provide much physical insight.
For many optical systems it is sufficient to analyze the paraxial performance.
The paraxial approximation assumes that the plane waves make small angles with respect to the z-axis which means that the $x$ and $y$ wave vector components are $k_x,k_y<<1$.
In this regime the propagation factor $k_z=\sqrt{n^2k^2 - k_{\perp}^2}\approx nk - \frac{k_{\perp}^2}{2nk}$ simplifying to:
\begin{equation}
\label{eq:angular_spectrum2}
\mathcal{E}(x, y, z') = e^{-ink(z'-z)} \mathcal{F}_{x,y}^{-1}\{ e^{-i\frac{k_{\perp}^2}{2nk}} \mathcal{F}_{x,y}\{\mathcal{E}(x,y,z)\}\}.
\end{equation}
The real space solution that is derived from this propagation equation is the classical Fresnel diffraction equation,
\begin{equation}
\label{eq:fresnel_diffraction}
\mathcal{E}(x, y, z') = \frac{e^{-ink(z'-z)}}{\lambda z} \iint_{-\infty}^{\infty} \mathcal{E}(u, v, z) e^{\frac{ik\left[(x - u)^2 + (y - v)^2\right]}{2 (z' - z)}} \mathrm{d}u \mathrm{d}v.
\end{equation}
The far-field approximation ($z \gg x, y$) and the substitution $\theta_x = \frac{x}{\lambda z}$, $\theta_y = \frac{y}{\lambda z}$ results in the Fraunhofer diffraction integral,
\begin{equation}
\label{eq:fraunhofer_diffraction}
\mathcal{E}(\theta_x, 
\theta_y) \propto \iint_{-\infty}^{\infty} \mathcal{E}(u, v) e^{-i2\pi(\theta_x u + \theta_y v)} \mathrm{d}u \mathrm{d}v.
\end{equation}
This shows that the far-field distribution of an electromagnetic wave is its Fourier transform.
A perfectly aligned lens, or the mirrors of a telescope, brings rays that come from infinity to the focal point.
This is identical to a far-field transform, and therefore the operation of a lens can be described by the Fraunhofer diffraction integral.

\begin{figure}[ht]
  \centering
  \script{plot_telescope_psfs.py}
  \includegraphics[width=1.0\linewidth]{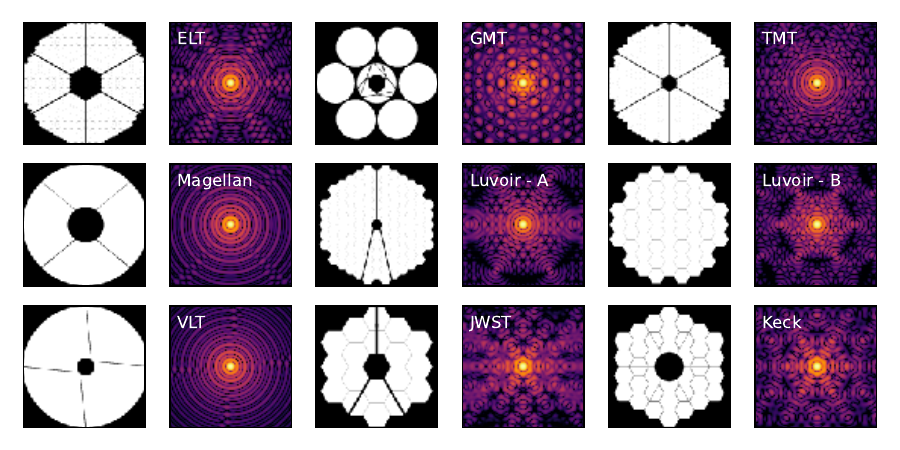}
  \caption{Telescope pupils and their \acp{psf} for several ground and space based telescopes.
  \acp{psf} have been normalised to the peak of the \ac{psf} and are on a logarithmic scale down to $10^{-6}$.}
  \label{fig:telpsfs}
\end{figure}

The \ac{psf} is defined as the response function of an optical system to a point source at infinity.
We show the \acp{psf} (defined as the square of the Fourier transform of the telescope pupil $\Pi$) equal to $|\mathcal{F}_{x,y}[\Pi]|^2$ for several telescope pupils seen in Figure~\ref{fig:telpsfs}.

\section{The Lyot Coronagraph}

Stellar coronagraphs have now been in use for several decades.
However, the first coronagraph was developed by Bernard Lyot to observe the corona of the Sun in the 1930s \citep{Lyot39}.
It took over half a century before astronomers applied coronagraphs to image the faint circumstellar environment by blocking starlight. 

\begin{figure}[ht]
  \centering
  \includegraphics[width=0.95\linewidth]{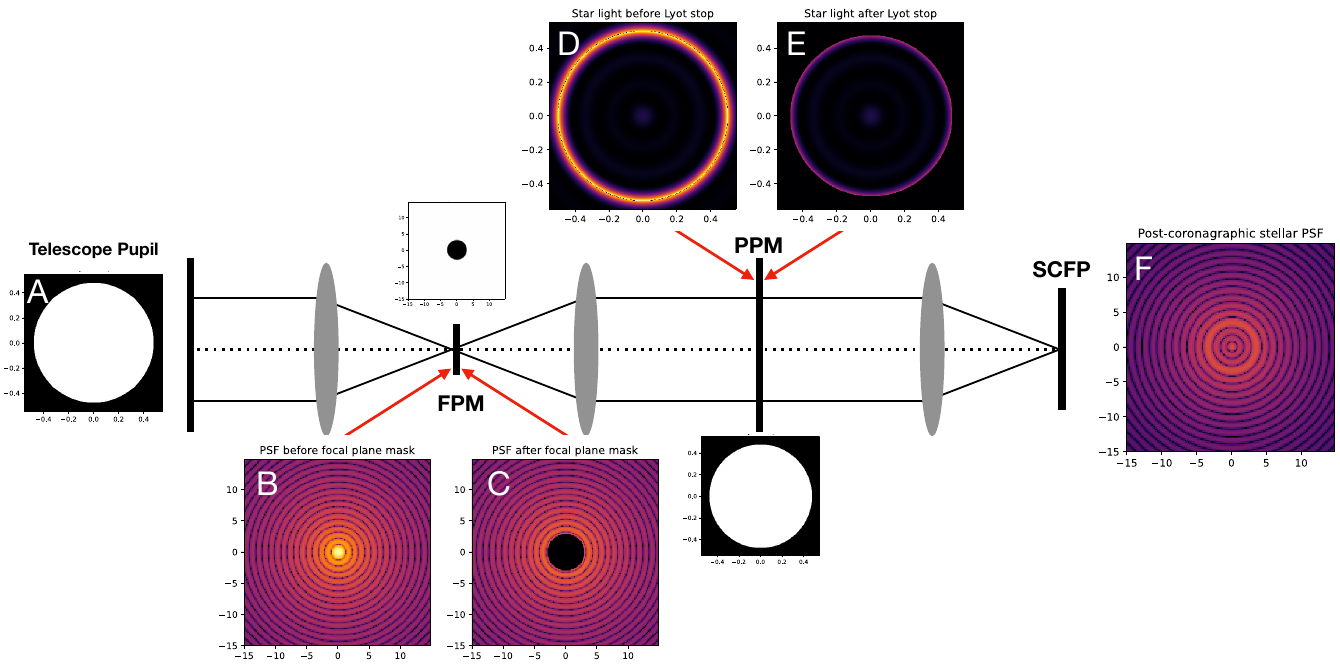}
  \caption{The Lyot coronagraph.
  A telescope with a circular unobstructed pupil (image A) points at a star along the optical axis of the telescope and instrument.
  The telescope pupil is reimaged through the instrument optics to a focal plane seen at B, showing an Airy core surrounded by diffraction rings.
  A \acl{fpm} then blocks the central region of the stellar image, resulting in the image seen at C.
  Another set of optics reimages C to a pupil plane at D.
  The removal of the Airy core at the \acl{fpm} redistributes light into a broad ring around the edge of the reimaged telescope pupil.
  The Lyot stop (more generally referred to as a \acl{ppm}) blocks this star light, at a cost of reduced off-axis throughput and decreased angular resolution.
  A final set of relay optics then form the \acl{scfp} at F.}
  \label{fig:lyot}
\end{figure}

The first coronagraph to successfully image a debris disk was a Lyot coronagraph built by \citet{Vilas87} and it imaged the edge-on circumstellar disk around Beta Pictoris in 1984 \citep{Smith84}.
The optical layout of the Lyot coronagraph can be generalised in Figure~\ref{fig:lyot}, with the letters A-F representing the images present at that location in the coronagraph light path.
The telescope pupil (A) is reimaged into a focal plane of the sky (B) where a \ac{fpm} that has high absorptivity and low reflectivity blocks the light from any on-axis source (C).
Optics then form an image of the resultant pupil (D) to an intermediate \ac{pp}, where a \ac{ppm} - the Lyot stop - is located.
The diffraction of starlight around the \ac{fpm} results in a ring of light around the diameter of the reimaged telescope pupil (D), and Lyot stop blocks this ring of light.
A second optical system then reimages this light onto the \ac{scfp} (F).
Any circumstellar objects outside the radius of the \ac{fpm} then pass through unimpeded through the coronagraph and are subsequently reimaged in the \ac{scfp}.
The light rays pass through the coronagraph optics to form an image at F with only minor modification: the reimaged pupil at D is superficially very similar to the telescope pupil A.

\begin{armarginnote}[]
\acc{fpm}
\acc{ppm}
\acc{scfp}
\end{armarginnote}

For an on-axis source, the removal of the Airy core plus attendant diffraction rings significantly modifies the wavefront passing through the coronagraph, resulting in a flux redistribution at D where the flux is concentrated in a ring whose peak brightness lies along the perimeter of the reimaged telescope pupil, extending both beyond the radius of the pupil and into the centre of the pupil.
The purpose of the Lyot stop is to remove as much of this ring of light as possible, whilst maximising the throughput of the pupil image D for off-axis sources.
Decreasing the diameter of the \ac{fpm} changes the \ac{fwhm} of the ring of light at D, which requires a smaller Lyot stop to block - but the throughput of the pupil for off-axis sources then decreases.
Decreasing the Lyot stop aperture has a second impact in that the reduced pupil diameter increases the \ac{fwhm} of the images in the final focal plane F, spreading the flux from the off-axis sources over a larger area in the detector and degrading the angular resolution of the telescope and instrument.
The optimal diameters of the \ac{fpm} and Lyot stop aperture are then driven by the science requirements - how close to the central star (measured in diffraction widths at the lower spatial resolution) should the coronagraph be able to transmit light from off-axis objects in the field of view.

\section{Parameters to optimize}

Clear apertures (ones with no secondary obscurations that block light from a simple circular aperture) can have solutions that perfectly remove any on-axis light.
However, real systems are not ideal and generally don't have a clear aperture.
Even more importantly, stars are not ideal point sources.
Solutions that only work for point sources are already out of the question then.
The current and future generation of coronagraphs are designed to take on non-ideal environments.
This means that the coronagraph is optimized for a specific list of parameters during the design process.

\begin{armarginnote}[]
\acc{iwa}
\acc{owa}
\end{armarginnote}

\begin{figure}[ht]
  \script{plot_dark_hole_geometries.py}
  \centering
  \includegraphics[width=1.0\linewidth]{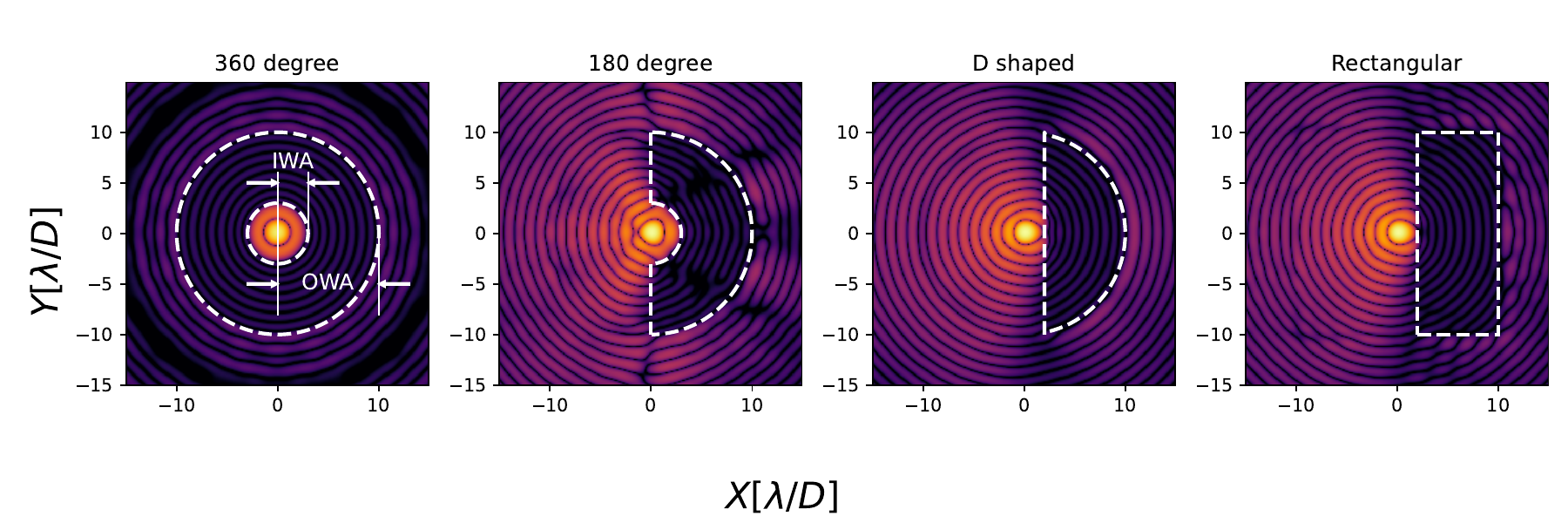}
  \caption{Geometries of dark holes commonly used in high contrast imaging.}
  \label{fig:coronagraph_focal_plane_definitions}
\end{figure}

The first set of parameters are the \ac{iwa} and the \ac{owa}.
The \ac{iwa} is the angular separation where the throughput is 50\% of the peak off-axis throughput, and the \ac{owa} is set by the design and optical properties of the coronagraph.
The \ac{iwa} and \ac{owa} set the smallest and largest angular separation where the coronagraph will suppress the stellar halo, producing a dark hole region in which to image planets and/or circumstellar material.
Some coronagraphs, such as the \ac{ovc} or \ac{fqpm}, only have an \ac{iwa}.
Figure~\ref{fig:coronagraph_focal_plane_definitions} shows the different types of coronagraphic dark hole geometries commonly used with the definition of the \ac{iwa} and \ac{owa}.

\begin{figure}[ht]
  \script{plot_planet_throughput.py}
  \centering
  \includegraphics[width=1.0\linewidth]{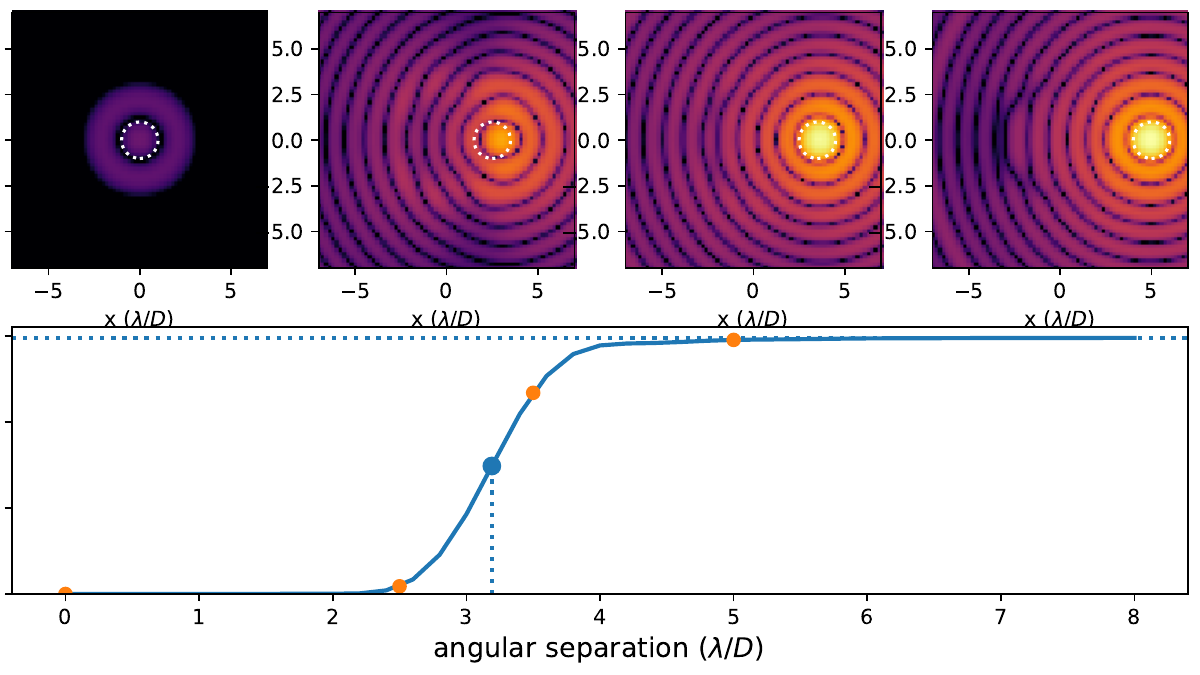}
  \caption{Definition of planet throughput, showing the integrated flux over a two \ld{} aperture with increasing distance from the star.
  The dotted line is the maximum throughput at an infinite off-axis angle.
  The vertical dashed line and dot show the IWA angle at 50\% throughput of the maximum throughput.
  The inset of the four post-coronagraphic images correspond to the four orange points on the throughput curve.
  The planet's post-coronagraphic image is strongly warped at/close to the edge of the focal plane mask.
  The presented simulations here correspond to the Apodized Lyot Coronagraph that is shown in Section~\ref{sec:nulled_lyot}.}
  \label{fig:planet_throughput}
\end{figure}

The contrast and throughput are the next set of parameters that we need to optimize for.
The contrast sets the amount of starlight that is left after the coronagraph.
It is important to define the term contrast as this can mean many different things.
\citet{ruane2018review} provide a thorough overview of the different metrics and their definition.
The contrast is defined as
\begin{equation}
C = \frac{\eta_*(\vec{r})}{\eta_p(\vec{r})}.
\end{equation}
Here $\eta_*(\vec{r})$ is the fractional throughput of the star at focal plane position $\vec{r}$ integrated over a photometric aperture.
This is then divided by $\eta_p(\vec{r})$ the fractional throughput of the planet in the same photometric aperture (see~Figure~\ref{fig:planet_throughput}).
This normalizes the contrast w.r.t. the throughput of the planet, which is important because the planet throughput usually varies as function of angular separation.
Both the contrast $C$ and $\eta_p(\vec{r})$ need to be included in the optimization process.
The first to make sure that the starlight is nulled and the second to make sure that the planet light is maintained.
This optimization has to be done over a certain spectral bandwidth $\Delta \lambda$.

The coronagraphs that are designed with only the previous set of optimization targets are not optimal in real environments.
In real instrument environments there are wavefront aberrations and small instrumental drifts.
These cause light to leak around the coronagraph and generate residual stellar {\bf speckles}, $\lambda/D$-sized spots forming granular patterns which are caused by diffraction and interference effects when coherent beams undergo random scattering.
In this case these speckles originate from residual turbulence or non-common path aberrations.
The coronagraphs must be made robust against low-order wavefront errors and other instrumental drifts.
Other more practical things to consider are the precision with which we can align an instrument.
For example, how well can the Lyot stop be aligned?
The performance of a coronagraph might be extremely sensitive to the Lyot stop alignment, which means that theoretically the coronagraph delivers the contrast but practically it will never reach it.
Therefore, alignment tolerancing must be included in the coronagraph design to make sure the target performance is achieved.

In this way, there are many other nuisance parameters that can be included.
However, the numerical optimization will take significantly longer if more parameters are included.
A good coronagraph designer will therefore make a trade-off between which parameters are required, good to have and not significant.

\section{Beyond Lyot with complex pupil and focal plane masks}

The Lyot coronagraph was designed at the time when it was still difficult to precisely manipulate the phase of light with optics.
With the advent of more advanced manufacturing capabilities, very precise phase control became possible.
This was a significant boost to the design toolbox for coronagraphs.
The major downside of the classic Lyot coronagraph is that only the light that falls on the opaque \ac{fpm} gets blocked.
Any light that is off-axis will pass through the system.
This holds not only for light that comes from off-axis sources but also for aberrations that causes light to end up outside of the \ac{fpm}.
A bigger mask will be able to block a larger fraction of the light and therefore achieve a deeper contrast.
However, if the mask is made larger the \ac{iwa} also becomes bigger and that means fewer planets will be accessible.
A smaller \ac{iwa} is crucial both for the extremely large ground-based telescopes and the space-based telescopes. 

We note that the coronagraphic designs in subsequent sections are not optimised but are shown for illustrative purposes.

\subsection{Focal plane phase mask coronagraphs}

Focal plane phase masks offer a solution by phase shifting a part of the \ac{psf} (usually the core of the \ac{psf}) which then leads to destructive interference at the Lyot plane.
The Lyot stop blocks the areas where the light does not destructively interfere.
The 50\% encircled energy radius is on the order of $\sim$\ld{} for almost all aperture shapes.
This means that it is possible to achieve perfect destructive interference with a mask that has a size on the order of \ld{}.
This is the central idea that was used to design the Roddier and Roddier (RR) phase mask \citep{roddier1997stellar}.
The RR mask covers the core of the Airy pattern that contains 50\% of the encircled energy and phase shifts the core by $\pi$ to cause destructive interference.
The RR mask works well for monochromatic light, but over broad spectral bandwidths it degrades in contrast.
Diffraction causes wavelength scaling of the \ac{psf} that either makes the \ac{fpm} too big or too small compared to the \ac{psf}.
Masks made out of multiple concentric rings, such as the dual-zone phase mask \citep{soummer2003achromatic}, were proposed to increase the spectral bandwidth.

\begin{armarginnote}[]
\acc{ovc}
\acc{fqpm}
\acc{agpm}
\end{armarginnote}

Further achromatization for larger spectral bandwidths is possible by designing inherently achromatic phase masks.
Inherent achromatic masks are scale invariant, which means that they always look the same regardless of the size of the \ac{psf}.
The \ac{fqpm} splits the focal plane into four quadrants and applies a checkerboard $0,\pi$ phase pattern \citep{Riaud00}.
The \ac{fqpm} was implemented on the \ac{sphere} instrument \citep{Boccaletti04} and is currently the coronagraph on \ac{jwst} with the smallest \ac{iwa} and has characterized a planet at 1.8 \ld{} \citep{franson2024jwst}. 

If the planet falls on one of the four transition lines between the phases, its transmission is significantly reduced.
Increasing the number of phase steps to the continuous limit results in a phase ramp about the optical axis, and this is called the \ac{ovc}.
The \ac{ovc} uses a phase mask with a vortex pattern where the phase changes with the azimuthal angle: $\phi=q \cdot \theta$ with $q$ the charge of the vortex (the number of times the phase wraps around) and $\theta$ the azimuth angle.
One implementation of an \ac{ovc} is the \acl{agpm} \citep[\acs{agpm}; ][]{Mawet05b}.
This uses subwavelength gratings to impart a charge 2 phase ramp in a \ac{fpm}.
The grating manufacture limits the \ac{agpm} to a charge 2 vortex, but higher charges are preferred to make a better match to the small but finite diameter of stellar disks for nearby stars.
A more general challenge is that the \ac{ovc} diffracts all the light out of the \ac{pp} only if the telescope pupil is unobstructed.
Secondary support structures and a centrally obscured pupil scatter significant amounts of light back into the \ac{pp}.
A suitable Lyot stop can block this stellar leakage, but at a cost of planet throughput.
The impact of the central obscuration can be mitigated with the use of a telescope pupil grey apodizer to make a \acl{ravc} \citep[\acs{ravc}; ][]{Mawet13a}.

Both phase masks completely null out all on-axis light from a clear aperture.
This property is why both the \ac{fqpm} and the \ac{ovc} have been extensively studied over the past several decades.
The \ac{iwa} of both coronagraph approaches 1 \ld{} which makes coronagraphy possible at the diffraction limit!

\subsection{Pupil plane mask coronagraphs}

All stars have a small but finite angular diameter on the sky, typically $\lambda/100$ or less but increasing to $\lambda/10$ for the closest stars.
In the visible, median stars for \ac{hwo} are going to be $\lambda/10$, with some a little larger, becoming even larger as the apertures for \acp{elt} will be larger by a factor of a few: Proxima Centauri has a diameter of 1 mas, which is $\frac{1}{3}$ to $\frac{1}{6}$ \ld{} for the \ac{elt} and the \ac{gmt}.
The disk of the star can be treated as a set of incoherent point sources, and so the small but finite angular size of the star means that the sensitivity of contrast of focal plane coronagraphs varies as a function of angle.
Given that the star is tens of thousands to millions of times brighter than the target planet, even a small amount of stellar leakage can overwhelm the flux from the planet.

Apodizing the telescope pupil provides an opportunity to redistribute the light in the \ac{scfp} to form `dark zones' around the target star where an exoplanet can be imaged.
Strictly this is not so much a coronagraph as a modification to the \ac{psf} of the instrument - all objects in the focal plane have the same \ac{psf}, both stars and exoplanets together.
As long as the angular diameter of the star is smaller than the Airy core of the \ac{psf}, \ac{pp} coronagraphs are not impacted by the diameter of the star or by residual tip tilt vibrations that are not removed by the \ac{ao} control loop, making them a robust alternative to the more efficient, smaller \ac{iwa} \ac{fp} coronagraphs. 
Suppressing diffraction in the \ac{psf} requires destructive interference using coherent light from the Airy core, but decreases the encircled energy of the core of the \ac{psf}.
Since the exoplanet \ac{psf} is identical to the stellar \ac{psf}, the planet throughput decreases too.

The earliest pupil apodizations (referred to as Shaped Pupil masks or Shaped Pupil Plates) were with binary amplitude masks \citep{Jacquinot64,Kasdin05} but these had low throughputs, a significant increase in the FWHM of the resultant \ac{psf} (impacting the encircled energy of the planet and the \ac{iwa}) and a narrow opening angle ($<45^\circ$) of the dark zone.
Improvements in the searching of the large dimensional space of possible solutions resulted in the improvement to throughputs of 50\%, working angles of 2.5 to 15 \ld{} and contrasts of $10^{-6}$ \citep{Carlotti11}.
Optimizations are also generalised so that they can be designed for arbitrary telescope pupils, so that the secondary obscuration, support structures, and even gaps between segmented mirrors can be accounted for and avoided.
Along with their achromatic performance, this makes pupil apodization suitable for \ac{elt} telescope pupils and enables dynamic coronagraphs using micromachine mirrors \citep{Leboulleux22b,Carlotti23} to account for a dynamically changing pupil (e.g. missing and swapped mirror segments).

\begin{armarginnote}[]
\acc{spp}
\acc{app}
\acc{gvapp}
\acc{vvc}
\end{armarginnote}

Another approach is to apodize in phase only, with the \ac{ppm} set by the geometry of the telescope pupil \citep{Codona04}; this was realised and subsequently demonstrated on-sky with the \acl{app} \citep[\acs{app}; ][]{Kenworthy07} which used variations in the thickness of a piece of diamond turned Zinc Selenide to impart a phase shift across the pupil with a central wavelength of 4 microns.
An \ac{app} built and installed in NAOS/CONICA VLT camera \citep{Kenworthy10} led to the first coronagraphic image of Beta~Pictoris~b \citep{Quanz10} and the direct imaging discovery of the exoplanet HD~100546~b \citep{Quanz13}.
The original algorithms found solutions with $180^\circ$ dark `D' shaped regions next to the star, but a more general theory to APP optimisation \citep{Por17} finds both 180 and 360 degree solutions and additionally finds solutions consisting of regions of integer multiples of $\pi/2$ radians.

The first \ac{app} optics were chromatic: the Optical Path Difference (OPD) was a function of the refractive index of the transmissive material, with the suppression decreasing with increasing bandwidth.
Achromatic phase shifts can be implemented using the principle of geometric phase: the vector-APP \citep[vAPP; ][]{Snik12} replaces the classical phase pattern $\phi_{\textrm{c}}(\vec{r}) = n(\lambda) \Delta d(\vec{r})$,
with the ``geometric phase'' \citep[known as the Pancharatnam-Berry phase; ][]{Pancharatnam,Berry}.
The vAPP phase pattern is imposed by a half-wave retarder with a patterned fast axis orientation $\theta(\vec{r})$.
The geometric phase is imprinted on incident beams decomposed according to circular polarization state: $\phi_{\textrm{g}}(\vec{r}) = \pm2\cdot\theta(\vec{r})$, with the sign depending on the circular polarization handedness.
As this fast axis orientation pattern does not vary as a function of wavelength (with the possible exception of an inconsequential offset/piston term), the geometric phase is strictly achromatic.
Vector-APP devices are produced by applying two liquid-crystal techniques: any desired phase pattern is applied onto a substrate glass through a \textit{direct-write procedure} \citep{directwrite} that applies the orientation pattern $\theta(\vec{r})$ by locally polymerizing the alignment layer material in the direction set by the controllable polarization of a scanning UV laser.
Consecutive layers of birefringent liquid-crystal are deposited on top of this alignment layer, which subsequently self-align \citep[``\textit{Multi-Twist Retarders}''; MTR ][]{MTR} with predetermined parameters (birefringence dispersion, thickness, nematic twist) to yield a linear retardance that is close to half-wave over the specified wavelength range.
Additional layers broaden the wavelength range to over an octave in wavelength, at a cost of an absorption feature due to the carbon-carbon bonds within the liquid crystal and glue layers.
The vector-APP devices required additional optics (typically a half-wave plate and Wollaston prism) to isolate the two circular polarizations and produce two separate \acp{psf} with dark holes on opposing sides of the central star \citep{Snik12}.
By adding a phase diffraction grating onto the APP phase pattern to make a grating vector APP \citep[gvAPP; ][]{Snik12,Otten14}, the other optics are no longer required.
The two coronagraphic \acp{psf} are separated diffracted into the $m=\pm 1$ order, with a $m=0$ ``leakage term'' non-coronagraphic \ac{psf} with flux of a few percent of the original star left in the undeviated beam, acting both as an astrometric and photometric reference \citep{Otten17,Sutlieff24}.
The grating effect means that the \ac{psf} centroids vary as a function of wavelength, and so the gvAPPs are ideal for imaging onto integral field units and image slicers \citep{Sutlieff21,Sutlieff23}.
This liquid-crystal technology has enabled coronagraphic designs that were previously impossible to manufacture, including the coronagraphic modal \ac{wfs} \citep{Wilby17}, sparse aperture masking with multiple holograms \citep{Doelman21}, complex amplitude Vector Vortex Coronagraphs \citep[VVC; ][]{Snik14}, and triple grating coronagraphs \citep{Doelman20} that redisperse the \acp{psf} back into white light coronagraphic \acp{psf} for VVCs \citep{Doelman23,Laginga24}. 
A comprehensive review of the coronagraphs enabled by the liquid crystal technology is given in \citet{Doelman2021a}.

\section{Apodized Lyot Coronagraphs}\label{sec:nulled_lyot}

The fundamental goal of any coronagraph is to block the light from the star, while passing through the light of the planet.
Any type of \ac{fpm} coronagraph, like the Classic Lyot coronagraph, will find that it is very difficult to completely null out the star.
This is due to a mismatch between the modes of the incoming electric field from the aperture and the \ac{fpm} filtered electric field.
The propagation through a Classic Lyot style coronagraph can be described by two propagations.
The first is a propagation through the area that is covered by the mask itself, $m_1$, and the second is the negative of the mask $m_2$.
Together, these two masks cover the full focal plane.
The propagation of the electric fields then follow as,
\begin{equation}
    \mathcal{E}_{\mathrm{out}} = t\invfourier{m_1 \fourier{\mathcal{E}_{in}}} + \invfourier{m_2 \fourier{\mathcal{E}_{in}}}.
\end{equation}
Now let's simplify this by substituting $\hat{m}=\fourier{m_1}$, $m_2 = 1 - m_1$ and using the Fourier convolution theorem. This results in a simplified equation,
\begin{equation}
    \mathcal{E}_{\mathrm{out}} = \mathcal{E}_{in} + (t-1)\hat{m}*\mathcal{E}_{in}.
\end{equation}
The output is the sum of two electric fields, the original input electric field and a filtered electric field.
The filtered light is allowed to diffract light outside the geometric pupil because this is blocked by the Lyot stop.
Therefore, the condition for perfect nulling is found by setting the output over the geometric pupil to zero. 
\begin{equation}
    \mathcal{E}_{in} + (t-1)\hat{m}*\mathcal{E}_{in} = 0.
\end{equation}
This condition can only be fulfilled if the incoming electric field is an eigenfunction of the filtering operator.
\begin{equation}
    \hat{m}*\mathcal{E}_{in} = \gamma \mathcal{E}_{in}.
\end{equation}
Then
\begin{equation}
    \mathcal{E}_{in} + (t-1)\gamma \mathcal{E}_{in} = 0,
\end{equation}
and a perfect null is achieved when $(t-1)\gamma = -1$.
The normal pupil illumination is uniform and is not an eigenfunction of the filtering operator.
Therefore, it is not possible to perfectly null the starlight with any type of Lyot-style \ac{fpm}.
These eigenfunctions are Prolate Spheroids and they generate theoretically perfect nulls if combined with the Roddier \& Roddier coronagraphs \citep{soummer2003stellar}.

The incoming pupil amplitude must therefore be apodized.
This insight led to the development of various coronagraphs with different combinations of phase/amplitude pupil apodization and opaque/phase shifting \acp{fpm}.

\subsection{The Apodized Pupil Lyot Coronagraph (APLC)}

\begin{figure}[ht]
  \centering
  \includegraphics[width=0.8\linewidth]{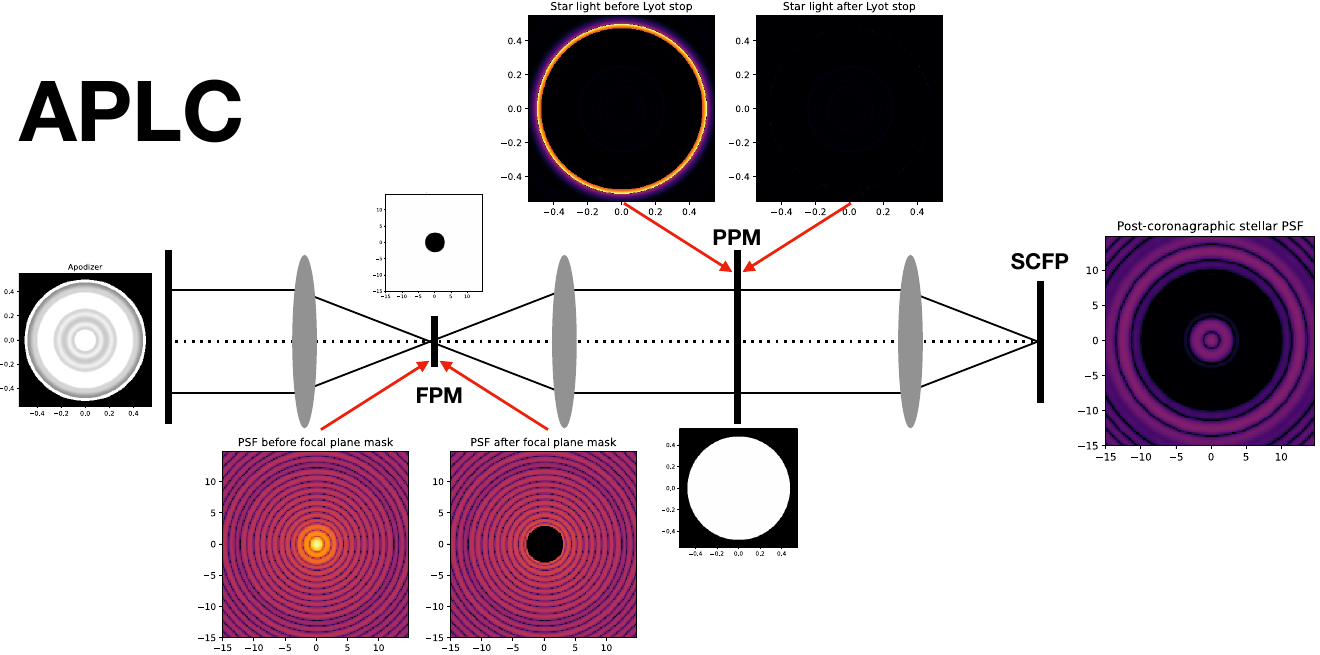}
  \includegraphics[width=0.8\linewidth]{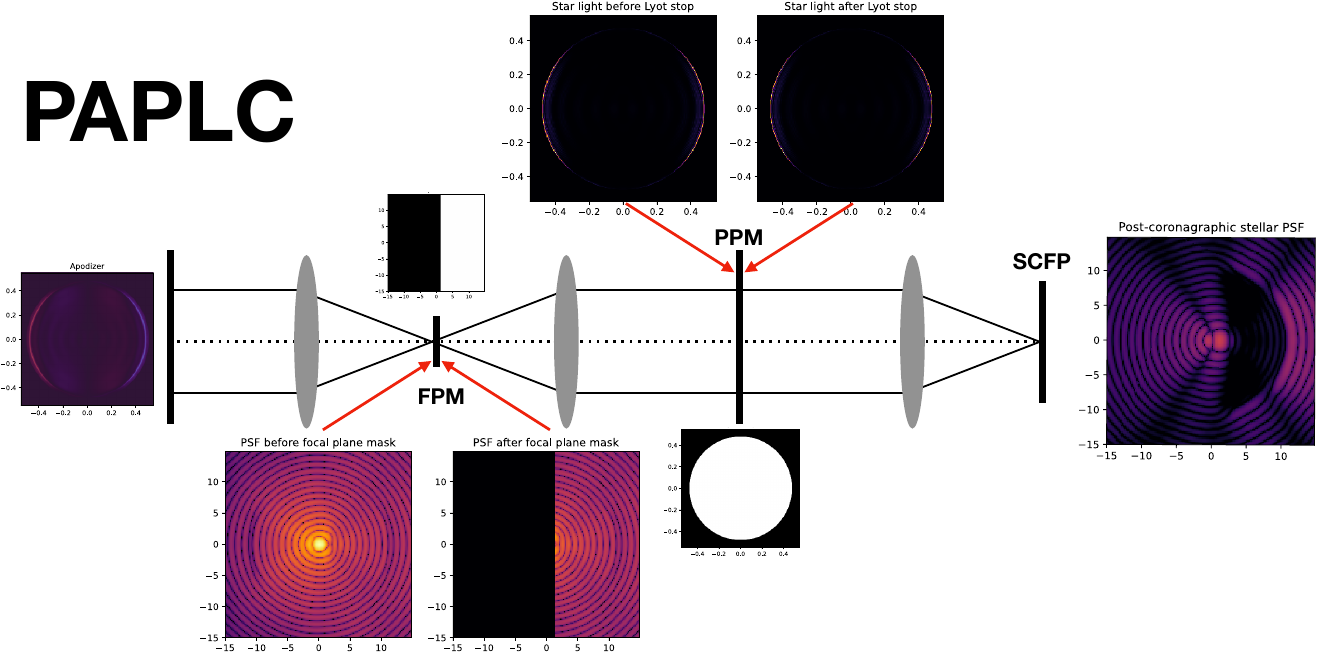}
  \caption{The APLC and PAPLC layouts, showing the pupil, focal plane images and masks}
  \label{fig:coro_aplc}
\end{figure}

The first modification of the Lyot coronagraph that people experimented with was the \ac{aplc}.
The \ac{aplc} uses achromatic grey-scale pupil apodizers to better match the entrance amplitude distribution to the \ac{fpm} (see Figure~\ref{fig:coro_aplc}).
These coronagraphs are quite robust and in use in many different high-contrast imaging instruments such as \ac{sphere} \citep{beuzit2019sphere}.
Currently, there is still active research in the optimization strategies for the \ac{aplc} such as for future segmented space telescopes \citep{zimmerman2016lyot} and for upgrades of ground-based instruments \citep{nickson2022aplc}.
The major downside of the \ac{aplc} is the lower transmission due to absorption in the apodizer.
A variation on the \ac{aplc} will be used by Coronagraph Instrument (CGI) on the Nancy Gracy Roman Space Telescope \citep{krist2023end}.
CGI uses a \ac{hlc} where the phase effects of the \ac{fpm} are taken into account during the design process.
The metallic coatings on the focal plane mask substrates are not completely opaque.
The mask will leak starlight at a certain level causing residual speckles.
For broad spectral bandwidth dark holes, the exact transmission and phase shift needs to be taken into account for the pupil apodization mask \citep{kuchner2002coronagraph}.
The \ac{hlc} will be the first coronagraph in space with active wavefront control for dark hole digging \citep{krist2023end}.

\subsection{The Phase Apodized Pupil Lyot Coronagraph (PAPLC)}
\label{sec:paplc}
Phase apodization instead of amplitude apodization is twice as effective in improving the performance of the Lyot Coronagraph \citep{Por20}.
This comes from the range of allowed apodization values.
Instead of being limited between $[0, 1]$, it is possible to use phase offsets to increase the range to $[-1, 1]$ and allows for complex value apodization.
The solutions for circular dark holes are found to be discretized; the phase is either 0 or $\pi$ \citep{Por20}.
This is reminiscent of the solutions found for optimal APPs \citep{Por17} and binary shaped pupils \citep{Carlotti11}.
Full circular dark holes are not very efficient since they require strong phase patterns to create the dark holes which causes significant Strehl loss \citep{Por17}: one-sided dark holes are much more efficient in terms of Strehl. 

The PAPLC is the baseline for the GMagAO-X \citep{Males24}, and has been implemented in the Space Coronagraph Optical Bench \citep[SCOoB; ][]{Ashcraft22,vanGorkom22} at Arizona.

\subsection{The Phase Induced Amplitude Apodization Complex Mask Coronagraph}

\begin{figure}[ht]
  \centering
  \includegraphics[width=1.0\linewidth]{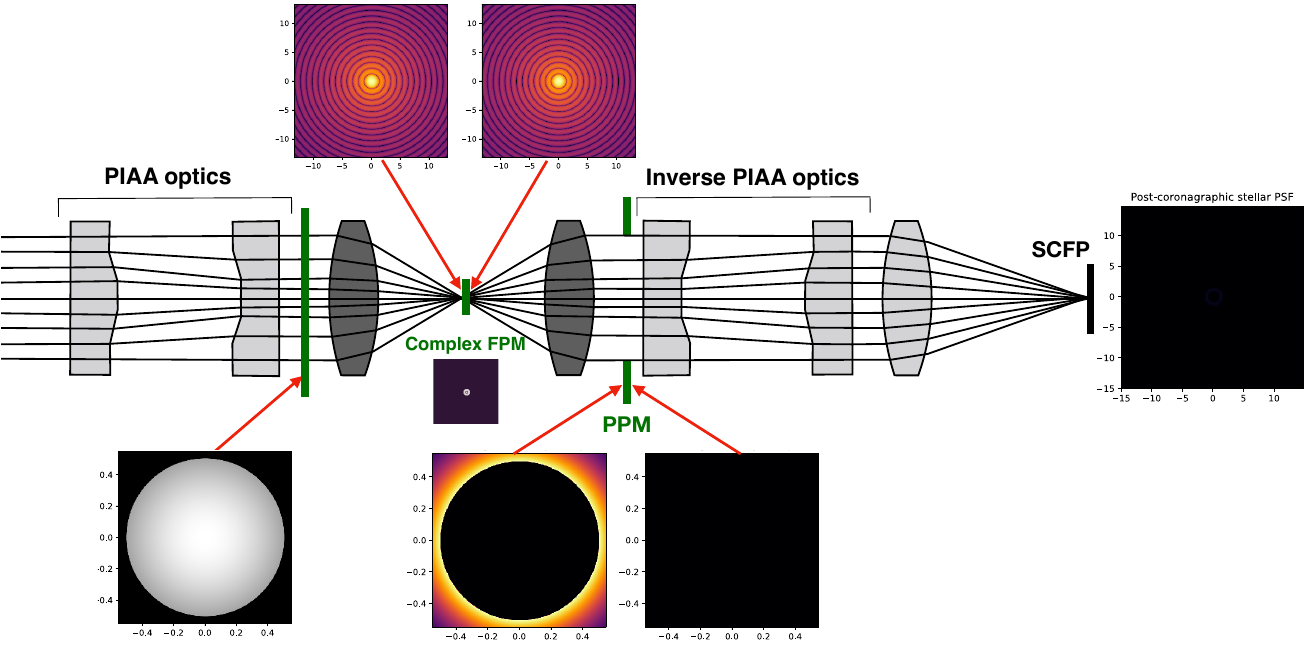}
  \caption{The PIAACMC coronagraph principle.}
  \label{fig:piaatypes}
\end{figure}

\ac{piaa} remaps the telescope pupil such that a star on the optical axis forms a \ac{psf} with no diffraction rings - typically a 2-D Gaussian profile \citep{Guyon03,Guyon05,Guyon14}.
The pupil remapping optics can be either transmissive or reflective, with reflecting optics more amenable to achromatization but more challenging to manufacture.
The optics induce aberrations for off-axis sources that are strong functions of increasing distance from the optical axis, significantly decreasing the Strehl ratio of these sources and lowering their effective sensitivity.
A reimaging system that reverses the optical aberrations of the first set of \ac{piaa} optics then reforms a final focal plane image with all off axis sources forming diffraction limited images.
An on-axis \ac{fpm} then blocks the starlight whilst allowing off-axis sources to propagate through to the \ac{scfp}.
The original design \ac{piaa} uses a hard edged apodizer, but by allowing the design to include other coronagraphs (an amplitude apodized Lyot coronagraph; AALC) or a complex mask coronagraph (CMC), they can approach the ideal coronagraph in their suppression - see Figure~\ref{fig:piaatypes}.

Original \ac{piaa} designs are for unobscured circular apertures: telescope pupils with secondary obscurations require reformatting of the pupil to make a continuous diffraction-free \ac{psf} in the coronagraphic focal plane.
The introduction of complex masks that can be manufactured to the required tolerances enable \ac{piaa}s for complex and segmented telescope pupils, suitable for space-based telescope designs such as the HabEx/LUVOIR concepts.
Results from the laboratory demonstration of a \ac{piaacmc} with a segmented aperture, \citet{Marx21} show contrasts of $3\times 10^{-8}$ from $4-9$\ld{} with 2\% bandwidth.
Most recently, the laboratory demonstration of high contrast with the \ac{piaacmc} coronagraph on an obstructed and segmented aperture \citep{Belikov22} shows $1.9\times 10^{-8}$ contrast achieved in a 10\% bandwidth between $3.5-8$\ld{}.
Ultimately the rejected light can form the basis for a \ac{wfs} to keep the \ac{piaa} pointed and aligned with the science target, and an integrated \ac{wfs} and coronagraph with \ac{piaacmc} has been demonstrated \citep{Haffert23a}.

\section{Spatial mode demultiplexing}

The principle of \ac{smd} is to use waveguides as modal filters on the complex electric field to separate them into different optical pathways.
An early example is the filtering of the input of a nulling interferometer.
Two subapertures from a wavefront are brought to a focus with a $\pi$ radian phase shift between them, forming a set of Young's fringes on the sky with an on-axis stellar source nulled out quadratically.
The electric field of the fringes are antisymmetric, with the complex electric field changing sign across the central null.
By using a single mode fiber at the focus  which only permits transmission of the lowest electromagnetic mode of the circular aperture of the fibre (HE$_{11}$), the sign change of the (point anti-symmetric) electric field results in a deep null, and wavefront aberrations are minimised \citet{Serabyn06,Haguenauer06}.
Any planet that is sitting in an adjacent transmissive region of the fringe pattern (and is on the face of the fibre) has a point symmetric electric field, and will couple into the fiber, albeit with a reduced transmission.

This principle is used to minimise the contribution of speckles in the focal plane at the location of the planet, where the Airy core of the planet is injected into a single mode fibre \citep{Mawet17}.
High coupling efficiency is possible \citep[with a theoretical maximum of 81\% for an Airy core into the Gaussian HE$_{11}$ mode; ][]{shaklan1988coupling} and reflected light around the fibre used to sense and minimise speckles whilst the incoherent light of the planet remains constant and injected into the fibre. 

The \ac{ovc} has a planet throughput of 0.7 at 1 \ld{} of the on-axis null \citep{Mawet05b}, and placing an optimally matched single mode fibre at the location of the null provides suppression of the star much greater than the transmission loss of the coronagraph for a planet at 0.5 to 1\ld{} from the star, resulting in a peak transmission of 0.2 at 0.9 \ld{}.
This principle is called \acl{vfn} \citep[\acs{vfn}; ][]{Ruane18} and different telescope pupils with both \ac{pp} and \ac{fp} \ac{ovc} are explored in \citet{Ruane19}, with on-sky results demonstrated in \citet{Echeverri24}.

\ac{vfn} provides almost no information on the azimuthal position of the planet, and because it is off-axis it does not couple with the highest efficiency.
One solution is to use a \acl{mspl} \citep[\acs{mspl}; ][]{LeonSaval13}.
This replaces a single-mode fiber with a cluster of fibres in a close-packed configuration that optimally match the focal plane with a \ac{ovc}.
The properties of the coupling increase the planet throughput and partial localisation of the planet \citep{Xin22}.
The \ac{vfn} and \ac{mspl} feed high dispersion spectrographs to perform High Spectral Dispersion High Contrast Imaging \citep[][ and this Review]{Snellen15}, where the speckle field changes slowly with wavelength and can be approximated as a constant background over limited wavelength ranges.

\ac{smd} can also suppress the diffraction halo of a star by using a single mode fibre that has a diffraction null crossing the face of the fibre.
The sign change in the electric field across the null means that a fibre centered on the null has a significantly reduced electric field propagated through the fibre, but the Airy core of the planet will couple with high efficiency.
This technique works in the narrow band, but since the diffraction halo scales radially with \ld{}, the null line no longer passes through the centre of a fibre that is offset from the optical axis defined by the star.
The effect can be made to work across a significantly wider bandwidth by having two successive nulls cross the fibre area: as the two nulls move out radially with wavelength, the two sign changes across each null compensate each other to first order.
An \ac{app} with modest apodization can be made to squeeze two null crossings closer to each other, and using a hexagonal lenslet array matched to the circular single mode fibres enables a high efficiency transmission of planet flux \citep{haffert2021fundamental} through to the spectrograph, but with a large bandwidth called \acl{scar} \citep[\acs{scar}; ][]{Por20a,Haffert20}.
This principle has been demonstrated in laboratory experiments with $360^\circ$ and $180^\circ$ dark regions from 0.8–2.4 \ld{} around the star \citep{Haffert20}.
In these experiments, the $360^\circ$ was designed for an unobscured telescope pupil and created a measured stellar null of $2-3 \times 10^{-4}$, and a $180^\circ$ SCAR was designed for a telescope aperture with central obscuration and spiders and reached a null of $1\times 10^{-4}$. 

\begin{figure}[ht]
\script{plot_smf_app.py}
  \centering
  \includegraphics[width=1.0\linewidth]{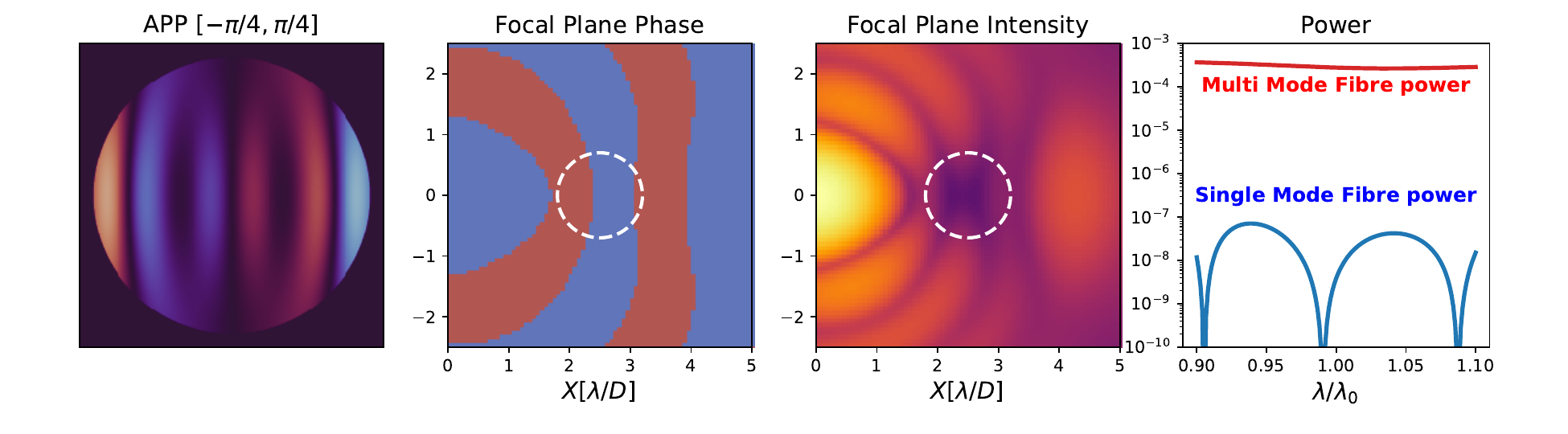}
  \caption{Demonstration of the SCAR.
  An \ac{app} modifies the distribution of the diffraction rings in the focal plane, moving two null crossing points to be within the same aperture (white dashed circles on the figures).
  A multi-mode fibre propagates all the energy within the circle, but for a single mode fibre the electric fields cancel out, resulting in a suppression of approximately $10^{-4}$ over a broad bandwidth around a central wavelength of $\lambda_0$.}
  \label{fig:scar}
\end{figure}

\subsection{Quantum optimal detection with SMD}

Quantum metrology is the field that uses quantum mechanics to enhance sensitivity and resolution in sensing devices.
The tools from quantum metrology can be used to derive fundamental measurement limits through the determination of the Quantum Cram\'er-Rao (QCR) bounds \citep{braunstein1994statistical}.
The limits on the variance of a measurement can be derived from the QCR bounds.
These principles were used to show that an optimal detection scheme can actually resolve incoherent point sources well within the classical Rayleigh diffraction limit \citep{tsang2016quantum}.
A SMD that sorts the incoming wavefronts into separate modal channels with photon counting detectors saturates the QCR bound.
Therefore, SMD is an optimal detection scheme for equal brightness point sources.

Most of the quantum optimal imaging research is focused on equal brightness point sources or low-contrast extended objects.
Exoplanet imaging is a situation with an extreme contrast ratio.
The optimal imaging limit could be different.
In \citet{deshler2024achieving}, the quantum limit for exoplanet imaging is derived.
And it is found that SMD with the telescope's eigenmode basis is quantum optimal. 
However, what is also found is that the \ac{piaacmc} and \ac{ovc} approach this fundamental limit for sub-diffraction limited separations within a factor of two.
Beyond separations of 1 \ld{}, the PIAACMC approaches the limit within 10\% making it nearly quantum optimal for resolved objects.
The photonic lanterns are devices that perform SMD albeit not in the eigenmode basis of the telescope.
Another avenue is to use Multi Plane Light Converters (MPLC).
The MPLC uses multiple phase plates that are separated by some distances to implement arbitrary unitary operations.
That also means that MPLCs could implement optimal SMDs \citep{deshler2024experimental}.

\section{Wavefront Sensing and Correction}

The coronagraph designs assume that the incoming wavefronts from all the astrophysical sources in the field of view are flat, and that the optics in the coronagraph are ideal, propagating and modifying these wavefronts without distortion to the final \ac{scfp}.
In reality there are several factors that cause deviations of the wavefronts from this ideal: (i) optical manufacturing limitations, (ii) environmental conditions (both static and dynamic) within the instrument and the telescope, and in the case of ground based telescopes, (iii) the wavefront residuals from the Earth's turbulent atmosphere partially corrected with a high order adaptive optics system.

\subsection{Adaptive Optics}

\begin{figure}[ht]
  \centering
  \includegraphics[width=1.0\linewidth]{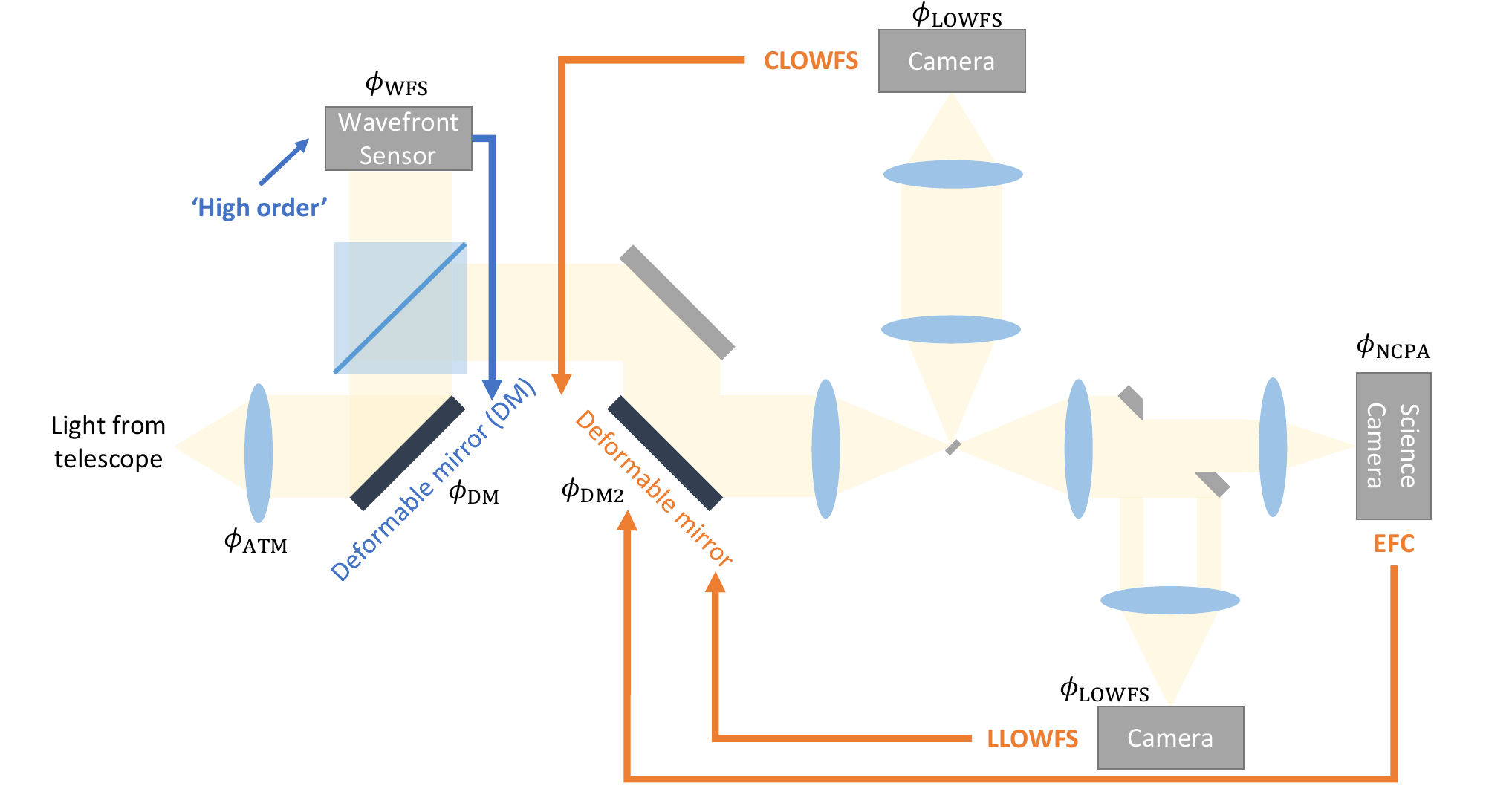}
  \caption{A schematic overview of an \ac{hci} instrument (GMagAO-X) showing the optical components, closed loops, and \acp{wfs}.
  Light from the telescope is corrected by a high order \ac{ao} system with a high order \ac{dm}, forming a closed loop that corrects a large fraction of the atmospheric turbulence.
  A beamsplitter or dichroic sends science light into a second AO system.
  This second \ac{dm} provides lower order \ac{ncpa} corrections via feedback from three possible sources - a CLOWFS, a LLOWFS and EFC from the \ac{scfp}.}
  \label{fig:aosystem}
\end{figure}

Adaptive optics sense the turbulence introduced by the Earth's atmosphere $\phi_{ATM}$ using Wavefront Sensors which measure a wavefront $\phi_{WFS}$, reconstruct an estimate of this turbulence and apply it to an electronically actuated deformable optical element - typically a \ac{dm}\footnote{There are several other optomechanical devices that exploit other optically active principles to modify a wavefront.} - within the instrument to modify the incoming turbulent wavefront and flatten it.
With the \ac{dm} upstream of the WFS, and an \ac{ao} computer providing the calculation of wavefront measured by the \ac{wfs} and applying this correction $\phi_{DM}$ to the \ac{dm}, this forms a {\bf closed loop}, where the response of the \acp{dm} correction is seen by the \ac{wfs} in the instrument, see the left-hand side of Figure~\ref{fig:aosystem}.

\begin{armarginnote}[]
\acc{dm}
\acc{wfs}
\acc{ao}
\end{armarginnote}

Incoming light is split using a dichroic or grey beamsplitter, sending some of the light to the science camera and the rest to the \ac{wfs} camera.
Many \ac{ao} systems take advantage of the fact that the optical path difference (OPD) introduced by the Earth's atmosphere above ground based telescopes is achromatic, despite OPD amplitudes of several tens of microns across large telescope apertures.
A consequence is that a wavefront measurement at a shorter wavelength (typically at optical or NIR wavelengths) will provide correction for all longer (science) wavelengths.
For 8m class telescopes looking at bright stars, when the wavefront sensing is carried out at the same wavelength as the science camera $(\lambda_{WFS}=\lambda_{SCI})$, the theoretical contrast is $10^{-6}-10^{-7}$ but when $\lambda_{WFS}$ is in the optical and $\lambda_{SCI}$ is in the infrared, the limit is set by the scintillation chromaticity induced by Fresnel propagation through the atmosphere to $10^{-4}-10^{-5}$ within one arcsecond \citep[see ][ for a discussion of the limits to \ac{ao} for HCI]{Guyon05-1}.

The Earth's atmosphere is highly dynamic and changes on a timescale of milliseconds, but the wavefront reconstruction and correction on the \ac{dm} is not instantaneous, leading to a small but significant time lag between measurement and the application of the correction.
One measure of the turbulence is the Fried length $r_0$ which is the radius of a circle where the mean variance at a wavelength $\lambda$ is equal to 1 rad$^2$.
The Fried length is a function of wavelength $r_0 \propto \lambda^{6/5}$ and is typically quoted at $\lambda = 0.5 \mu m$.
Adaptive optics is a complex and mature field in its own right, covering atmospheric turbulence, optomechanics, engineering control theory, wavefront sensing and information theory (each of these topics would be a review in their own right), but for now we refer the reader to \citet{Guyon18} for reviews on these topics. 

\begin{figure}[ht]
\script{plot_ao_residuals.py}
  \centering
  \includegraphics[width=1.0\linewidth]{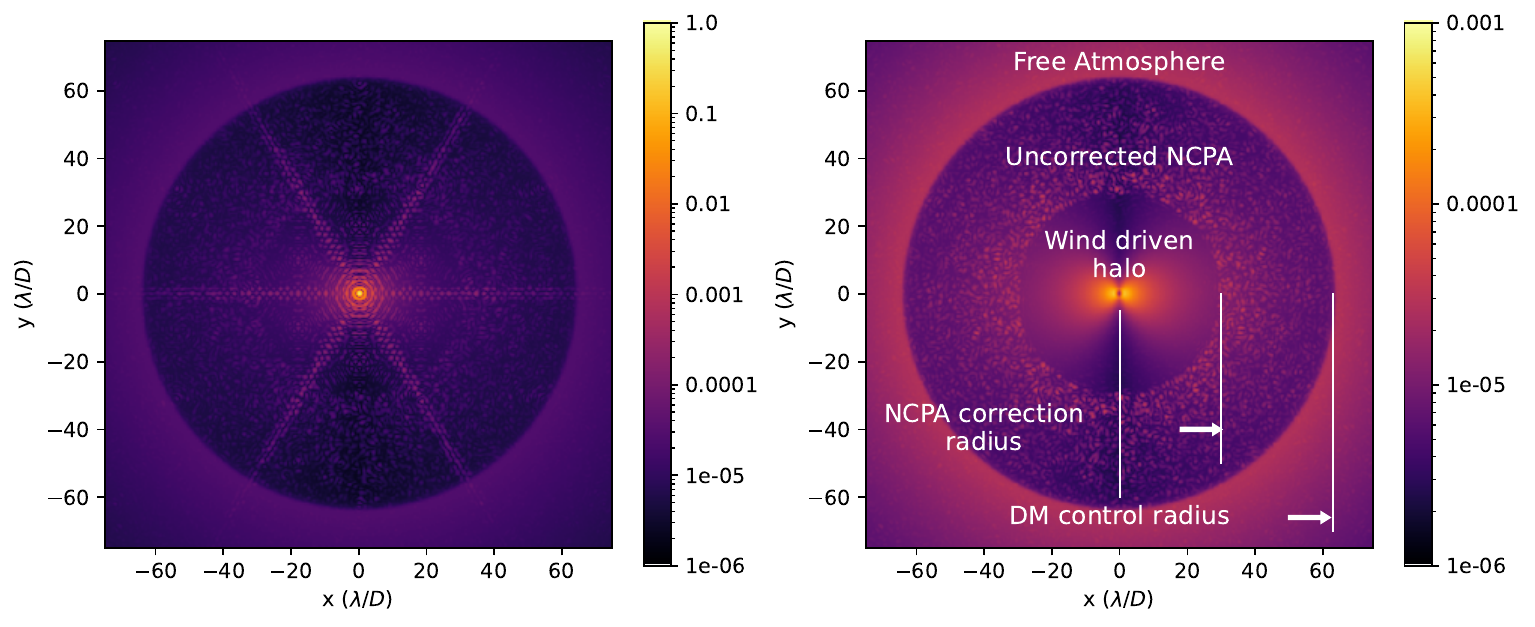}
  \caption{Closed loop of an ELT \ac{ao} system with the PSF (left) and ideal coronagraphic PSF (right).
  In the coronagraphic \ac{psf}, the outermost radius is due to the \ac{dm} control bandwidth, with the free atmosphere outside this radius.
  Residual speckles are seen in this annulus, down to the \ac{ncpa} correction radius.
  Close to the star the \ac{psf} is dominated by the wind driven halo.
  }
  \label{fig:aopsf}
\end{figure}

We simulate an ELT \ac{ao} system feeding a high contrast instrument that contains an ideal coronagraph: the \ac{dm} has 128 actuators across its diameter, resulting in a DM control halo 128 diffraction widths in diameter (see Figure~\ref{fig:aopsf}). 
The system speed is 2kHz, with \ac{wfs} observing at 0.5 microns and the science camera wavelength of 800 nanometres through an atmosphere with $r_0=0.16$m and wind speed of 37 m/s (exaggerated to emphasise the wind driven halo).
A second 60 by 60 actuator DM provides \ac{ncpa} correction.

In Figure~\ref{fig:aopsf}, a plume of speckles is seen around the central image, a result of the unsensed atmosphere along the leading edge of the telescope crossing into the pupil before it is corrected by the \ac{ao} system. 
Instantaneous speckles average out over time to an azimuthally symmetric halo, with an extended wind-driven halo that can be time varying in orientation and even asymmetric in the presence of strong turbulence \citep{Cantalloube18}.


\subsection{Non-Common Path Aberrations}

Aberrations can be sensed and corrected to the point of the last \ac{wfs} in the optical path in the high contrast instrument.
Ideally the sensed wavefront $\phi_{WFS}$ is identical to the wavefront delivered to the science camera $\phi_{SCI}$, but since the wavefront is divided at the beamsplitter, aberrations introduced in the two separate optical paths will result in differences between the \ac{wfs} wavefront and the Science camera wavefront.
The differential aberrations between this \ac{wfs} and the final \ac{scfp} are referred to as \acp{ncpa}.
These are characterised by using the Pearson coefficient $\rho(t)$, which quantifies the changes in intensity in the speckle field in the \ac{scfp} between two images taken a time $t$ seconds apart.
The Pearson coefficient is zero for completely uncorrelated intensities between the two images, and unity means that the speckle field is static between the two images.
Studies using the internal calibration source of the SPHERE instrument \citep{Martinez12,Milli18,Vigan22} and SCExAO internal sources and on-sky data \citep{Goebel18} show three characteristic timescales of decorrelation.
The shortest timescale is on the order of the speed of the closed AO loop, and is attributed to the evolving atmosphere above the telescope aperture (about 2ms for the 1.68 kHz SCExAO system).
The next two timescales seen in on-sky SCExAO system are expressed as $\rho(t)=0.018 e^{-t/0.40}-5.6\times 10^{-5}t + C$, where an exponential decay of a few seconds is added on a linear decrease over several minutes.
Both effects were noted with on-sky and internal calibration source measurements (with different timescales for the two cases).
Similar effects and timescales were seen in the analysis of \acp{ncpa} in SPHERE \citep{Vigan22}, where the shorter timescale has an amplitude of a few nanometers and is attributed to fast internal turbulence within the enclosure generated by heat from actuation motors.
The second is a slow quasi-linear decorrelation on the order of a few $10^{-3}$ nm rms s$^{-1}$ that acts on timescales from minutes to hours and is due to changing optical figures on the internal optics caused by temperature gradients and other sources of mechanical flexure.

First generation HC instruments provided little to no active \ac{ncpa} measurement and in-situ mitigation strategies, but the presence of \acp{ncpa} were a significant impact on the sensitivity of these instruments at smaller \ac{iwa}, reducing the predicted contrast from their designs.
Subsequent instruments have taken multiple approaches to \ac{ncpa} at every stage of the instrument's life cycle.

During the design of the instrument, minimising the number of optics that can contribute to the \ac{ncpa} \citep[by making the optics optomechanically and thermally stable; ][]{Absil24}.
During operation of the instrument, methods for estimating the wavefront at the final \ac{scfp} and providing active feedback to minimise the speckle field in the dark hole.
After the data is obtained, algorithms that provide estimates of the science camera \acp{psf} at all times during the observations can then minimise and/or remove the speckle field in port processing.

\section{Challenges for Segmented Telescopes}
Monolithic mirror telescopes are ultimately limited by their transport from manufacturing point to the observatory location, and so segmented telescope designs are used for diameters greater than 8m on the ground.
The Extremely Large Telescope projects are the European ELT, the Thirty Meter Telescope and the Giant Magellan Telescope.
All three telescopes have altitude/azimuth mounts, with segmented primary mirrors that have support structures holding a secondary mirror in front of the primary, blocking the central part of the telescope pupil.
The telescope pupils are shown in Figure~\ref{fig:telpsfs}.
The large apertures mean that the \ac{iwa} are on the order of 10mas for H band imaging, enabling direct imaging searches and characterisation, and enable upgrade paths and new instruments to be built based on the experiences of the first generation instruments.
For the ELT, all three first light instruments have \ac{hci} modes: METIS \citep{Brandl22,Absil24}, MICADO \citep{Sturm24,Huby24} and HARMONI \citep{Thatte22,Houlle21} that include coronagraphs mentioned earlier in this Review.
The challenges of atmospheric correction due to the wind driven halo, atmospheric dispersion, and the water vapour content of the Earth's atmosphere mean that ground based telescopes are mostly limited to \ac{hz} planets around nearby M dwarfs.

{\bf Missing/tilted segments:} Segmented mirror telescopes provide a challenge in that they require periodic cleaning, resulting in a varying transmission across the telescope pupil, and occasionally segments that are removed entirely for realuminization.
For the ELT, a baseline of 3 to 8 segments will not be available in the telescope pupil, changing nightly according to the realuminization schedule.

{\bf Atmospheric dispersion (ground based only):} The wavelength-dependent differential refraction introduced by the Earth's atmosphere is called atmospheric dispersion, which increases $\propto \sec(z)$ where $z$ is the zenith distance to an astronomical target.
In units of diffraction widths, the atmospheric dispersion gets larger for larger diameter telescopes, making it a challenge for \acp{elt} to observe science targets far from the zenith \citep{Kendrew08,Skemer09,van2020quantification}.
High order atmospheric dispersion correctors are required to produce diffraction limited imaging over wide bandwidths \citep{Kopon13}.
For the mid-infrared \ac{elt} instrument METIS, there is an additional complication due to the non-linear and variable nature of the atmospheric dispersion around the water bands, which make atmospheric dispersion correction far more challenging \citep{Absil22}.

{\bf Low Wind Effect: } When the wind speed within large telescope domes drop below a 3 m/s, low order large amplitude wavefront distortions are seen in the science camera \acp{psf} that are not sensed or removed by the \acp{wfs}.
This was initially discovered and characterised on \ac{sphere} \citep{Sauvage16}.
The most probable explanation are air temperature gradients formed next to the secondary support structure, whose temperature is anomalously deviant from the night time air temperature.
These temperature gradients then form piston-like aberrations within each sector of the telescope pupil, which the Shack-Hartmann \ac{wfs} is insensitive to detecting.
Discussions on mitigating it by adjusting the instrument design are described in \citet{Milli18} and control solutions for the Shack-Hartmann \ac{wfs} are presented in \citet{pourre2022low}.
Fast low order algorithms that are able to sense these modes can then provide feedback to the \ac{ao} system to remove this effect, such as Fast and Furious \citep{Wilby18}, and several other mitigation strategies with have been tested and verified on sky with the SCExAO/VAMPIRES system \citep{Vievard19}.

{\bf Petal modes: } With \ac{elt} telescopes, the large secondary mirror units require large secondary support structures whose projected thickness as seen on the telescope pupil can be several Fried lengths wide.
This creates a discontinuity in the wavefront reconstruction due to the spiders.
They fragment the pupil into unsensed separate petals because the thickness of the secondary support is several times greater than $r_0$.

Even monolithic mirrors have this problem too with thick enough secondary support structures, known together with the LWE as the island effect.
Differing approaches include apodizing each giant mirror segment individually \citep[Redundant Apodized Pupils; RAP ][]{Leboulleux22,Leboulleux22a} and  measuring the effect with a spatially filtered unmodulated pyramid \ac{wfs} \citep{Levraud24} or using the Fast and Furious algorithm  \citep[demonstrated on Subaru/SCExAO in][]{Bos20}.

The \ac{gmt} design differs in having seven large mirrors, each 8m in diameter.
Circular mirrors are arranged in a hexagonal pattern, and the large gaps between the edges of the mirrors leads to differential piston errors between the mirrors.
The current phasing sensor for the GMT is a pyramid \ac{wfs} \citep{Quiros-Pacheco22}.
However, the pyramid \ac{wfs} can only sense differential piston within half a wave and so larger signals get phase wrapped.
Atmospheric turbulence also reduces the pyramid \ac{wfs} sensitivity \citep{Bertrou-Cantou22}.
To solve these issues, GMT will use a two-stage system.
The pyramid \ac{wfs} for fine phasing and the Holographic Dispersed Fringe Sensor (HDFS) for coarse phasing of the differential piston between the seven segments \citet{Haffert22}.
A phase plate is placed in a \ac{pp} of the instrument, which produces radially dispersed fringes that encode the relative phase between pairs of \ac{gmt} mirrors.
The sensor has a dynamic range of 30 microns and can measure relative piston differences of 10 nm r.m.s.
The two-stage approach has been validated through simulations and lab tests on the High Contrast Adaptive optics Testbed (HCAT) \citep{Hedglen22, quiros2024giant}.

 {\bf Segment phasing: }
The ELT and TMT have another phasing challenge: the primary mirrors are made up of several hundred segments.
Each telescope has therefore developed its own dedicated phasing sensor.
The TMT's phasing sub-system is similar to Keck's.
The Keck 10m telescopes relied on active control of their primary mirror segments using both metrology from sensors between adjacent segments and a modified Shack-Hartmann \ac{wfs} that looked at the \ac{psf} formed from apertures straddling two adjacent mirror segments.
The original phasing algorithm for Keck mirror segments was able to go from 30 microns piston down to 30 nm in \citet{Chanan98,Chanan00}.
However, the original approach was not able to routinely reach 30 nm rms.
This impacts the performance of \ac{hci} instruments on Keck.
A very promising phasing approach is to use the \ac{zwfs} that demonstrated with phasing to 46nm precision on-sky \citep{vanKooten22}, and \citet{Salama24} using a Vector \ac{zwfs} which increased the Strehl ratio by typically 10\%.

The \ac{elt} has selected the ZEUS sensor \citep{Dohlen06}, which is also a \ac{zwfs} style sensor but then adjusted to work in seeing-limited conditions.
This approach has been demonstrated on testbeds \citep{Pfrommer18} and on-sky \citet{Gonte09} to reach the required levels.

\subsection{Specific challenges for space telescopes}

The advantages for telescopes and coronagraphs in space are immediately obvious: the turbulence, dispersion and transmission of the Earth's atmosphere no longer limit the achievable contrast, but the mirror sizes are limited by the rocket farings and their capacity.
Monolithic mirrors are possible up to a diameter of 2 to 3 metres, whose diffraction limit drives coronagraphs with the highest throughput and smallest possible \ac{iwa} so that the largest number of star systems can be imaged compared to an equivalent diameter segmented mirror.
Designs based around the \ac{ovc} that include additional hybrid elements (see \ac{aplc}) can produce exoplanet yields of around ten to twenty, assuming an unobstructed telescope aperture \citep{Stark24}.
Larger aperture telescopes can be realised with segmented mirror designs which can be folded into the faring of rockets and subsequently deployed, but have a more stringent requirement on mirror segment alignment due to the higher contrasts required - precisions of picometres are necessary to achieve $10^{-10}$ contrasts.
However, the alignment of the segments in these designs are susceptible to temporal drifts, leading to the generation of aberrations in the focal plane that are in the scientific region of interest.
Wavefront sensing and subsequent correction of these aberrations is therefore an important part of high contrast imaging.
Algorithms such as COFFEE have been demonstrated for the \ac{jwst} segmented primary pupil geometry \citep{Leboulleux20}.
The deployment of the \ac{jwst} mirror has shown drifts of 9.0 nm RMS per 48 hours \citep{Lajoie23}, demonstrating the need for active correction to obtain picometre precision for \ac{hwo} \citep{Laginja22}.
These present separate challenges for alignment and stability of the mirrors within the frame.
Target stars are too faint for wavefront measurement and alignment of the mirror structure, so the telescope must tune up on a much brighter star before slewing to the science target.
Ultra-stable structures are therefore required to keep the structure rigid during the slewing manoeuvre.
The key component level technologies have matured to a level where this is now feasible \citep{Coyle21}.

When the observations of target stars begin, there is an issue of what to do if drifts in the optics start to fill in the dark hole - it is better to actively maintain the dark hole at a cost of signal to noise whilst sensing is carried out, or is it better to let the drifts occur over the duration of the observation \citep{Pogorelyuk19,Redmond20}.

\section{Polarization effects} 

The field of high-contrast imaging is always hammering away at one noise floor after another.
It started with the common phase aberrations that dominate at low to moderate contrast.
After that, amplitude aberrations started to limit the contrast. This was solved by using multiple \acp{dm}.
Now, the contrasts that are achieved on-sky and on testbeds reveal another limit; polarization \citep{Schmid18,millar2022polarization,vanHolstein23, baudoz2024polarization}.
Polarization is an often underappreciated property of light.
The derivation shown in Section~\ref{sec:maxwell} actually also ignores the effects of polarization which was done for mathematical clarity.
However, light consists of two orthogonal polarizations states that do not interfere with each other.
That means that there are, at any time, always two beams of light propagating through our coronagraph that might interact in a different way with the instrument.
A more detailed treatment of polarization and physical optics propagation can be found in the literature \citep{McLeod14}.

\begin{figure}[ht]
  \centering
  \includegraphics[width=1.0\linewidth]{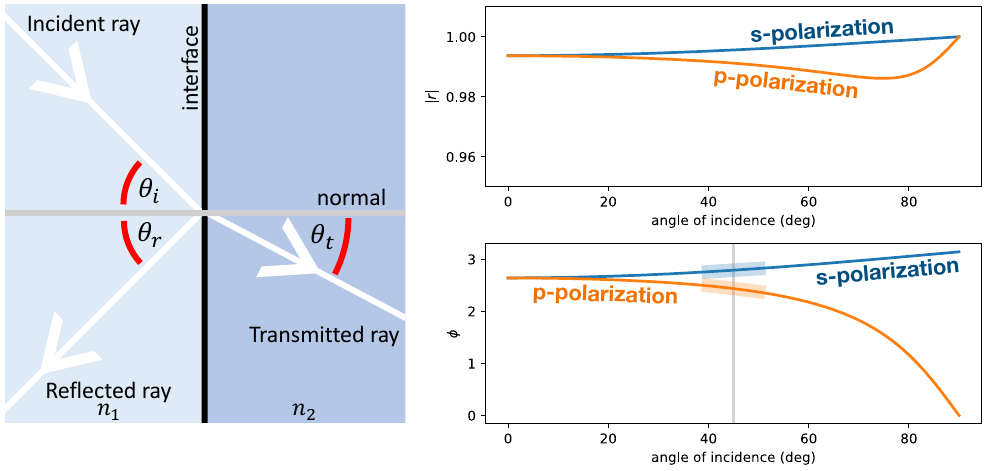}
  \caption{Reflectance and transmittance at an optical interface, and the differences between s- and p-polarization reflections at the interface.}
  \label{fig:fresnel_equations}
\end{figure}

The Fresnel equations describe how light is either reflected off or transmitted through an interface.
The definition of all variables for the incidence, reflected and transmitted wave are shown in Figure \ref{fig:fresnel_equations}. The Fresnel equations for plane wave interfaces are,
\begin{align}
r_s = \frac{n_1\cos{\theta_i} - n_2\cos{\theta_t}}{n_1\cos{\theta_i} + n_2\cos{\theta_t}},\\
t_s = \frac{2n_1 \cos{\theta_i}}{n_1\cos{\theta_i} + n_2\cos{\theta_t}},\\
r_p = \frac{n_2\cos{\theta_i} - n_1\cos{\theta_t}}{n_2\cos{\theta_i} + n_1\cos{\theta_t}},\\
t_p = \frac{2n_1 \cos{\theta_i}}{n_2\cos{\theta_i} + n_1\cos{\theta_t}}
\end{align}
Here, $r_x$ and $t_x$ are the reflected and transmitted amplitudes for polarization state $x$.
The Fresnel equations depends on the angle of incidence $\theta_i$.
While the transmitted angle $\theta_t$ is part of the Fresnel equations, it depends on the incoming angle through Snell's law, $n_2 \sin \theta_t = n_1 \sin \theta_i$.
Therefore, the Fresnel equations are a function only of the material and the incoming angle.
The equations show that different polarization states have different coefficients.
This wouldn't be a problem if the Fresnel equations did not also depended on the angle of incidence.
Any finite-sized beam, which means any physical plausible beam, has an angular spread because it will consists of a linear combination of plane waves.
Each of these waves will reflect of the interface slightly differently due to the angle of incidence depends.
This effect creates so called polarization aberrations \citep{Chipman89, Breckinridge15}.

Figure~\ref{fig:fresnel_equations} shows the phase change when reflecting off a silver mirror.
At 45 degrees incidence angle there is a substantial difference between s- and p-polarization, that becomes even larger at larger angles of incidence.
And, more importantly, a difference in the slope.
This means that a finite sized beam with a certain angular width will have different phase tilt aberration depending on the polarization state that enters the system.
In literature this effect is called the Goos-Hanchen shift.
In high-contrast imaging systems such effects create beam-shifts that limit the coronagraphic performance \citep{Schmid18, millar2022polarization}.

Polarization aberrations can be estimated using polarization ray tracing where the Fresnel equations are applied for every surface that is encountered during the raytrace \citep{ashcraft2023poke}.
A convenient way to represent the aberrations is in the Jones pupil format.
The Jones matrix is determined for each pixel in the pupil, which means that we end up with four pupil images (`xx', `xy', `yx' and `yy').
The Jones pupil can then be used in high-contrast imaging physical optics simulations to estimate the interaction of the polarization aberrations with coronagraphs \citep{Anche23} or adaptive optics residuals \citep{millar2022polarization}.
Recent simulations suggest that polarization aberrations could impact the extremely large telescopes at 1 to 3 \ld{} \citep{Anche23}.
However, the strongest effects are seen in space-based coronagraphic systems that require a deep raw contrast of $10^{-10}$ to $10^{-8}$.
Current systems typically encounter polarization aberrations at the $10^{-8}$ level \citep{mawet2011recent, seo2019testbed, baudoz2024polarization}.
Common strategies are to put the whole instrument between polarizers to ensure only one polarization state is propagated.
This ensures that the aberrations are controllable.
However, this approach loses 50\% of the light. Current research is focused on finding solutions to mitigate the effects, either by optimizing coatings of the optics \citep{balasubramanian2005polarization,miller2022birefringent}, or by active wavefront control \citep{mendillo2021dual} and by improving the polarization leakage of the coronagraph \citep{Doelman20, Doelman23}.

\section{Focal plane wavefront sensing}

Imperfections in the manufacture of the optics within a high contrast instrument and the changing environmental conditions result in speckles in the final \ac{scfp}.
Several techniques for optical sensing of these residual aberrations using telemetry or metrology within the instrument have been partially successful in sensing and removing these aberrations with closed loops, using actively deformable optics to provide correction for the sensed modes.
Ultimately, these methods cannot sense the time-varying aberrations within the last optical elements before the \ac{scfp}, and so several methods have been developed to measure and characterise optical aberrations using the images from the \ac{scfp}.
The fundamental challenge is that the vast majority of the science focal plane detectors are photodetectors, and so they do not record the complex amplitude of the incoming electric field in the wavefront, but record only the intensity.

The result is that an intensity image of the \ac{psf} cannot be uniquely inverted to give the phase and amplitude of the wavefront in the pupil of the system: an arbitrary wavefront can be represented as the weighted sum of a series of even $f(r)=-f(r)$ and odd $f(r)=-f(-r)$ point symmetric functions.
Odd functions produce \acp{psf} with point symmetry (e.g. tip/tilt, coma) but even functions produce the same \ac{psf} with the same amplitude regardless of sign - consider a wavefront with focus, which has $\psi(r) = a\sqrt{3}(2r^2-1)$, and both $a$ and $-a$ will result in the same intensity distribution in the focal plane.
Phase retrieval directly from the \ac{psf} is therefore an under constrained inverse problem.
One of the earliest methods for phase retrieval is an iterative method, the Gerchberg-Saxton algorithm (GS) \citep{Gerchberg72}.
The GS algorithm starts by picking a random phase for each pixel in the pupil and then it iterates between the focal plane and pupil plane where it replaces the amplitude either with the square root of the measured \ac{psf} or the known  pupil function.
By constantly iterating it solves the non-linear phase retrieval problem.
However, the GS algorithm can not fundamentally solve the sign degeneracy problem of even modes.
What happens during the iterations is that it uses the non-linear cross-talk from various modes to estimate the sign of the even modes.
Therefore, GS will not be able to measure the sign of defocus for example if there is only a defocus wavefront error.

In order to measure the complex amplitude of the \ac{psf}, a diversity must be introduced into the focal plane image, either temporally or spatially \citep[see ][ for a review of these]{Fienup13,Gonsalves14}.
A natural diversity is the introduction of a focus offset.
This can be easily implemented either by mounting a camera on a controllable stage or by using a deformable mirror \citep{VanGorkom21DMs}.
The classic phase diversity algorithm only works with normal \acp{psf} imaging.
The COronagraphic Focal-plane wave-Front Estimation for Exoplanets \citep[COFFEE ;][]{Sauvage12,Paul13, herscovici2018experimental} algorithm lifts this requirement by estimating the aberrations in a coronagraphic instrument.
Other types of diversity are also possible to implement, such as direct modifications of the pupil by adding obstructions \citep{martinache2013asymmetric, brooks2016polarization, bos2019focal, gerard2023high}.
Algorithms like phase diversity or COFFEE try to solve the full non-linear inverse problem.
However, this is a time consuming problem and during actual observations the wavefront errors might evolve faster than the required computing time.
Many variations of phase diversity are in development to work at real-time frame rates. 

One such method is the ``fast and furious'' (F\&F) algorithm \citep{Keller12} that has been verified in lab \citep{Wilby18} demonstrated on-sky \citep{Bos20} on the SCExAO system.
The F\&F algorithm is a modified GS algorithm that uses the previous measurement and control command.
This allows F\&F to solve the degeneracies while it's in closed-loop operation.

Other methods try to linearize phase diversity in either imaging mode or coronagraphy mode because linear methods allow for short computational times.
There is the Linearized focal plane technique \citep[LIFT; ][]{meimon2010lift}, or the linearized Analytical Phase Diversity \citep[LAPD; ][]{vievard2020cophasing}.
Certain coronagraphs also use algorithms that are specific to their behavior such as QACITS \citep{huby2017sky} for the \ac{ovc}.
QACITS is a tip/tilt centering algorithm to keep the star on top of the center of the \ac{ovc} mask resulting in a pointing stability of 2.4 mas over several hours. 
Most phase diversity methods aim to control only the low-order Zernike modes.
However, most coronagraphic dark holes are dug at medium spatial frequencies.
Linearized Dark Field Control \citep[LDFC;][]{miller2017spatial} is a method that extends beyond just a few low-order modes.
This method has been shown on-sky with the vAPP coronagraph and is being investigated to work with arbitrary coronagraphs \citep{miller2019spatial,ahn2023combining}.
The push for linear methods was due to computational complexity of the non-linear reconstruction algorithms.
Recent advances in machine learning have also made it possible to run deep neural networks with short inference times at real time rates.
These, often data-driven, methods can be used to solve the non-linear phase retrieval methods.
Variations on the neural network architecture allow for either normal imaging \citep{orban2021focal, zhang2021phasegan}, post-coronagraphic imaging \citep{quesnel2022deep} or used within an iterative scheme like `fast and furious' \citep{bottom2023sequential}.

All the phase retrieval algorithms are used to estimate the wavefront error and apply a correction, and so this loop is repeated until it converges to a flat wavefront.

\subsection{Dark hole digging}

A flat wavefront would ideally result in a deep and dark coronagraphic image.
However, coronagraphic dark holes are not completely free of starlight after wavefront flattening due to the existence of amplitude aberrations and higher-order frequency folding effects.
Instead of aiming to flatten the wavefront, \ac{hci} should aim to remove as much light as possible from the dark hole region.
This is informally referred to as `dark hole digging' \citep{malbet1995high, borde2006high}.

There are several control algorithms that dig dark holes.
There is the Energy minimization \citep{borde2006high}, speckle nulling \citep{martinache2012speckle, martinache2014sky}, \acl{efc} \citep[\acs{efc}; ][]{Giveon09} and stroke minimization \citep{pueyo2009optimal}.
The major difference between them is the exact cost function that is optimized.
The \ac{hci} algorithms were presented with a unified formalism by \citet{Giveon09,Giveon10} under \ac{efc}.
The fundamental concept for \ac{efc} is that it is possible to create speckles with the \ac{dm} that exactly cancel out the stellar speckles by destructive interference.
The \ac{dm} commands are found by minimizing the electric field within the dark hole region $S$,

\begin{equation}
    \hat{v} = \argmin_v \int_S |\mathcal{E}_{s}(x,y) + \mathcal{E}_{\mathrm{DM}}(x, y, v)|^2\mathrm{d}A.
\end{equation}

Here $v$ is the \ac{dm} command, $\mathcal{E}_{s}$ the stellar speckles in the coronagraphic focal plane and $\mathcal{E}_{\mathrm{DM}}$ the speckles created by the \ac{dm}.
Effectively, \ac{efc} finds the command that injects speckles with the exact same amplitude but opposite phase in the focal plane.
\ac{efc} is currently the most used and best optimized algorithm on testbeds and instruments \citep{Mennesson24}.

The electric field in the focal plane must be measured first before it can be cancelled.
Therefore, all dark hole digging algorithms are combined with an electric field sensing approach.
There are currently two main approaches; either pair-wise probing (PWP) where the deformable mirror is used to introduce diversity or the \ac{scc} that uses starlight that has been rejected by the coronagraph to create interference fringes.
The former method uses temporal diversity while the latter method uses spatial diversity.
Originally, \ac{efc} was presented in combination with PWP \citep{Giveon09}.

\begin{figure}[ht]
    \script{plot_efc_darkhole.py}
  \centering
  \includegraphics[width=1.0\linewidth]{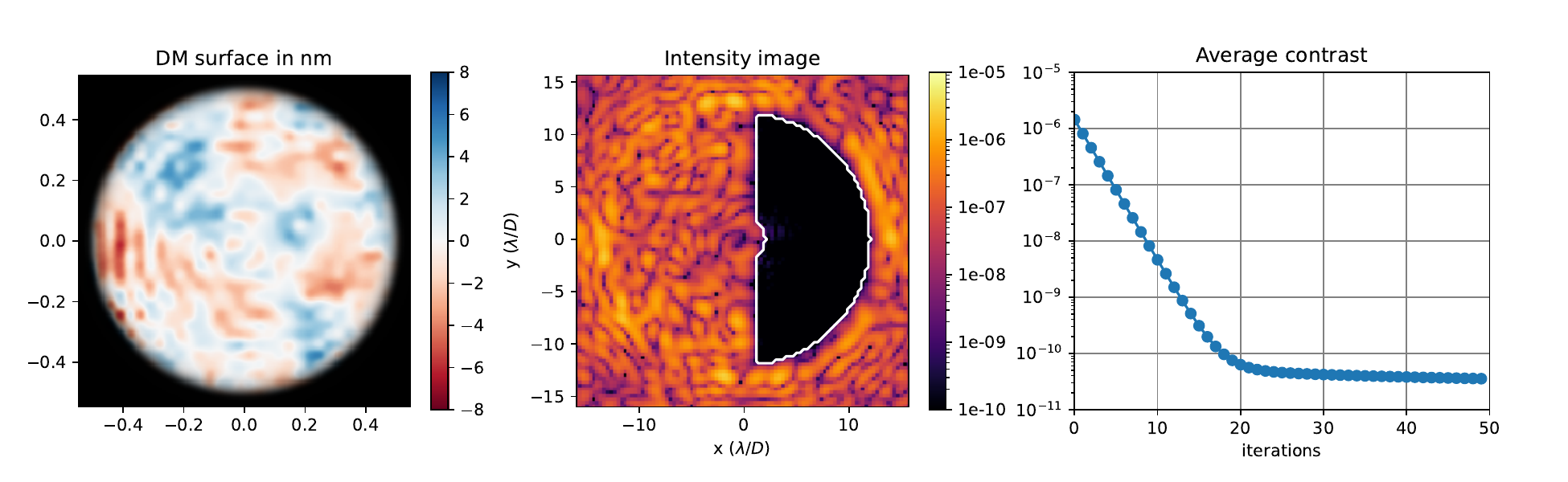}
  \caption{\ac{efc} generation of a dark hole.
  The dark hole is iteratively generated using the \ac{efc} algorithm.
The final \ac{dm} surface is shown in the left panel.
The final \ac{scfp} image with the target dark hole region is outlined with a white line.
Right hand panel: the average contrast in the dark hole as a function of iteration number.}
  \label{fig:efc_darkhole}
\end{figure}

The deformable element within the instrument can introduce phase shifts into the pupil image that then change the resultant complex amplitude for each location in the focal plane of the science camera.
For a single point source, all the light in the focal plane is coherent with respect to the core of the \ac{psf}.
Each pixel in the focal plane then becomes an intensity interferometer, and if four phase shifts that reasonably sample between $0$ and $2\pi$ radians are introduced into the instrument, then the four recorded \ac{psf} intensity images can be fit to give an estimate of the complex amplitude at each location in the \ac{scfp}.
It was shown that two pairs of complementary \ac{dm} actuations can provide enough phase diversity to measure the complex amplitude across the focal plane out to the spatial sampling of the mirror actuators.

The \acl{scc} \citep[\acs{scc}; ][]{Baudoz06} takes light from the telescope, splits it into two beams, filter one beam through a pinhole and recombine the two beams in a Fizeau configuration in the \ac{scfp}.
All the speckles in the \ac{scfp} are subsequently modulated by a set of cosine fringes, the relative position of the fringes encoding the complex amplitude of the electric field of the \ac{psf}.
Any exoplanets or other point sources emit photons that are incoherent with the speckles, and so fringes do not appear at the location of the planet.
Simulations demonstrate \citep{Galicher10} bandwidths of up to 5\% and contrasts possible down to $10^{-10}$.

There are currently two flavours of control in the literature.
The first is a model-free approach where the interaction of the \ac{dm} with the speckles is empirically calibrated, called implicit \ac{efc} \citep{haffert2023implicit}, and the second, more traditional, approach is where the interaction is determined from a numerical model of the optical system.
Every combination of control (\ac{efc} or iEFC) and sensor (PWP or SCC) lead to a similar performance in dark hole contrast \citep{desai2024comparative}.

One \ac{dm} enables correction of phase only aberrations in the science camera \ac{psf}.
If the aberrations include amplitude as well as phase, then correction with one \ac{dm} will result in one side of the \ac{psf} becoming dark, but the other side will not.
This is solved with two \acp{dm} (with one \ac{dm} between a focal and \ac{pp} to allow for both amplitude and phase control) which can act in tandem to correct both amplitude and phase aberrations and correct a 360 degree region in the field of view \citep{pueyo2009optimal}.

\ac{efc} has been tested on the internal sources of \ac{sphere} \citep{Potier20}, SCExAO \citep{ahn2023combining} and MagAO-X \citep{haffert2023implicit}.
Each of these instruments achieved a contrast of $\sim 10^{-7}$ at 100 to 200 mas from the optical axis in a few minutes.
The on-sky implementation took longer due to increased complexity by the atmospheric turbulence.
The first on-sky demonstration of \ac{efc} was with the \ac{aplc} coronagraph and \ac{sphere} \citep{Potier20,Potier22}, using an iterative scheme to clear a dark hole on one side of the science camera \ac{psf} down to an intensity of $10^{-6}$ an improvement of a factor of a few from 250 to 500 mas separations, and a second technical run with the \ac{fqpm} coronagraph in \ac{sphere} \citep{Galicher24}.
Additionally, implicit EFC \citep[iEFC; ][]{haffert2023implicit} has now also been demonstrated on-sky \citep{haffert2024sky, kueny2024magao}. %
A similar performance gain of a factor of a few has been achieved with iEFC on MagAO-X.

These algorithms can now clear out dark holes in coronagraphic images, limited by the wind driven halo from the AO system which can be filtered out using a low pass filter or post-processing algorithms.

\subsection{Modifications of the SCC for ground-based telescopes}

The downside of the \ac{scc} is that it uses a pinhole far away from the edge of the pupil edge.
The amount of light that a coronagraph then diffracts into the pinhole is very low.
The \ac{ncpa} and wind-driven halo in ground-based instruments are typically on the order of $10^{-6}$ to $10^{-3}$ which is significantly larger than amount of light that passes through the pinhole, resulting in a low fringe contrast for the classic \ac{scc}.
The low fringe contrast also implies a low SNR.
Several modifications to send more light to the pinhole used to generate fringes in the \ac{scfp} have been proposed to make the \ac{scc} more suitable for ground-based telescopes.
The first modification is a modified \ac{fpm}.
Instead of blocking light with a classic absorbing Lyot coronagraph \ac{fpm}, the light is steered away with the use of a phase ramp.
The ramp then centers the light on top of the pinhole and the throughput can be further improved by adding small amount of power in the \ac{fpm} within 3 \ld{}.
This approach is called the \acl{fast} \citep[\acs{fast}; ][]{Gerard18} and it increases the throughput by a hundred to a thousand fold which enables shorter exposures to capture the speckles in the \ac{scfp} and provide correction.
Laboratory measurements have demonstrated $3\times 10^{-4}$ contrast \citep{Gerard22}.
A similar approach can also be used to create a FAST-style PIAACMC coronagraph \citep{Haffert23a}.
Almost all coronagraphs diffract most of the star light close to the pupil edges that then falls off rapidly.
Moving the pinhole of the \ac{scc} closer in to the pupil edge will increase the throughput by a similar factor of a hundred to a thousand.
The downside is that the pinhole needs to be modulated to disentangle the fringes from the stellar speckles.
Several approaches to modulation have been proposed; temporal modulation \citep{martinez2019fast}, polarization modulation \citep{bos2021polarization} or spectral modulation \citep{Haffert22a}.

\subsection{Probing the electric field with the atmosphere}
A closed loop AO system leaves time-varying wavefront aberrations propagating through to the \ac{scfp}.
Short exposure images freeze these speckles and show that for a given location in the \ac{scfp} the flux changes as a function of time.
For thermal infrared science cameras on ground based telescopes, the science images saturate rapidly due to the thermal background of the Earth's atmosphere, and so the rapidly changing speckles naturally provide a ``free'' source of phase diversity.
Under the assumption that the \ac{ncpa} is changing on a much slower timescale than the speckles generated by the wind driven halo, the \ac{wfs} can provide an estimate of the complex amplitude of the \ac{scfp}, plus a fixed \ac{ncpa} term.
Phase Sorting Interferometry \citep[PSI; ][]{Codona13} enables a virtual interferometer to be constructed for each location in the \ac{scfp}, and the \ac{ncpa} calculated.
This was successfully demonstrated as a post-processing technique at thermal infrared wavelengths at the MMT Observatory with the Clio camera.

It is possible to go even a step further and estimate the electric field in each frame using wavefront sensor telemetry.
This allows for both speckle estimation and the detection of incoherent (with respect to the star) sources \citep{rodack2021millisecond, frazin2021millisecond}. 
This approach, while computationally intensive, would enable a duty cycle of 100\% while also probing for incoherent planet light at the same time.

\section{Rejected light wavefront sensing}

Focal plane wavefront sensing requires starlight to sense wavefront errors.
However, after digging a dark hole the starlight is gone which means there is no starlight anymore to sense wavefront errors.
Current research is investigating how to modify coronagraphs and algorithms to better utilize the light that has been rejected by the coronagraph to filter starlight.
Further more, measuring the centroid of the star (or other low-order aberrations) is not possible in the coronagraph \ac{scfp} due to the insensitivity to low order aberrations.
Low order aberrations can, however, be measured from the light rejected from the \ac{fpm}.

The Lyot-style coronagraphs that use opaque \ac{fpm} can be naturally extended to include low-order wavefront sensing capability by making the \ac{fpm} reflective - see Figure~\ref{fig:aosystem}.
The reflected light is reimaged onto a separate detector.
The encircled energy at a radius of 1$\lambda/D$ already contains 2/3rds of all the incoming star light.
The rejected light from a larger focal plane mask (2 to 3 \ld{}) will therefore contain almost all starlight.
This approach is called \citep[Coronagraphic low order \ac{wfs}; CLOWFS - ][]{Guyon09}.
The large amount of photons enable high-speed control and stabilization of the low-order modes shows that it can keep tip tilt to around $10^{-3}$\ld{} for a baseline telescope and observation of a 6th magnitude star \citep{Guyon09}.

For coronagraphs that use a phase mask in the focal plane (such as a Vortex Mask), another method is required.
By putting in a reflective Lyot stop, the rejected light from the pupil plane is reimaged to a separate camera and this forms the error signal for low order aberration measurements and is called the Lyot-based Low Order \ac{wfs} \citep[LLOWFS; ][]{Singh14,Singh15}.
The LLOWFS can measure tip-tilt down to $\sim 10^{-2}$\ld{} (equivalent to 2–12 nm at 1.6 \mum{}) per mode on the \ac{fqpm}, with on-sky results being somewhat larger \citep{Singh15}.
Both these methods deliberately introduce defocus into the rejected light image, so that the focus ambiguity is removed and tip-tilt and focus modes can be simultaneously measured.

A space telescope mission such as the Habitable World Observatory (HWO), which is based on the HabEx and LUVOIR studies, requires picometer stability to achieve the $10^{-10}$ contrast requirement.
This can only be achieved through active control but that also means that the wavefront errors must be measured with picometer level precision.
The segmented geometry of the HWO concept challenges the low-order wavefront sensors because the segments create high-spatial frequency errors.
The rejected light from the coronagraph masks or Lyot stops do not always containing enough diversity to recover the wavefront error signals such as segment misalignment.

The sensing is significantly improved by adding a \ac{zwfs}. The \ac{zwfs} is a common path interferometer that interferes a $\pi/2$ phase shifted core of the Airy pattern with the rest of the light.
The phase dimple typically has a diameter of 1-2\ld{}.
The resultant pupil image intensity directly encodes the phase of the wavefront.
This makes it ideal to recover wavefront errors at all spatial frequencies that are passed by the mask.
The \ac{zwfs} measurements are fed into the control loop for the \ac{dm} in the coronagraph to stabilize the wavefront - this was demonstrated for low and mid spatial frequencies with a \ac{zwfs} up to the \ac{dm} control radius in \citet{Ruane20}.

A modification of the \ac{hlc} uses a Dual Purpose Mask (DPM) for the \ac{fpm}, making a Dual Purpose Lyot coronagraph (DPLC).
This DPM is a tiered metallic focal plane occultor that suppresses starlight in the transmitted coronagraph channel, and a dichroic-coated substrate to reflect out-of-band light to a wavefront sensing camera.
It acts as a \ac{zwfs} in reflection, sending out-of-band light to a CLOWFS to maintain high contrast in the science focal plane \citep{Ruane23}.
A similar concept was tested on the HiCAT testbed where the light that is rejected from the focal plane mask was reimaged onto a \ac{zwfs} \citep{Pourcelot22,Pourcelot23}.
The \ac{zwfs} was able to control and reject the low-order wavefront errors that slowly creeped into the HiCAT dark zone and kept it clean of starlight \citep{Soummer22}.

By putting a \ac{zwfs} at the location of the \ac{fpm} in a PIAA coronagraph, this can approach fundamental sensitivity limits within the instrument \citep{Haffert23}.

\section{High contrast instruments for ELTs}

For the ELT, the first light instruments are METIS \citep{Brandl22}, MICADO \citep{Sturm24} and HARMONI \citep{Thatte22}, and they all have high contrast modes associated with them.
MICADO will have three Classical Lyot Coronagraphs, a \ac{gvapp}, and two sparse aperture masking modes \citep{Huby24}. 
METIS has been designed with \ac{hci} in mind \citep{Kenworthy16,Absil24} with a classical Lyot Coronagraph, \ac{ovc} designs with beamswitching capabilities and that use a \ac{ravc} design to accommodate the large secondary mirror in the ELT pupil, and \ac{gvapp} coronagraphs for both direct imaging $(2.9-5.3\mu m)$ and a high spectral resolution $(R\sim 100,000)$ IFS.
It is unique in being the only imager and spectrograph working beyond 3 microns out to 19 microns on an \ac{elt}.
HARMONI has a \ac{hci} mode \citep{Houlle21} with a \ac{spp} based on binary masks \citep{Carlotti23} and a dedicated ZELDA \ac{wfs} at 1.175 $\mu m$ to measure and correct \acp{ncpa} in the system.
The bands are $H$ and $K$ band and use an IFU with a fixed elevation Atmospheric Dispersion Corrector.
In the second generation of ELT instruments there is the Planetary Camera and Spectrograph \citep[PCS; ][]{Kasper21} with a goal of $10^{-8}$ at 15 mas angular separation from the star and $10^{-9}$ at 100 mas and beyond through the use of an extreme \ac{ao} system. 

The GMT	has the \ac{hci} GMagAO-X \citep{Males24} with a science requirement of photometry at a signal-to-noise ratio of 5 on a point source with a flux ratio of $10^{-7}$ or better with respect to its host star in a 10\% bandwidth filter at 4 \ld{}, working from 0.6-1.9 $\mu m$.
The baseline coronagraph design is a \ac{paplc}, and a stretch goal uses \ac{piaacmc} and transmissive complex focal plane masks.
Focal-plane low-order WFS (FLOWFS) and Lyot-plane LOWFS (LLOWFS) will make use of light rejected by the coronagraphs.

The Planetary Systems Imager \citep[PSI; ][]{Fitzgerald22} is a proposed instrument suite for the Thirty Meter Telescope (TMT).
PSI is optimized for high contrast exoplanet science from 0.5 to 13$\mu m$ and has a near-IR AO system feeding other systems, notably PSI-Red (2-5 $\mu m$), PSI-Blue (0.5-1.8 $\mu m$), and PSI-10 (8-13 $\mu m$) subsystems. 
The PSI-Red system would have coronagraphs that include the \ac{gvapp}, \ac{vvc} with a Lyot stop, and an \ac{spp} \citep{Jensen-Clem21}.

\section{Photonic versus bulk optics}

Optics change the complex amplitudes of wavefronts as they propagate through coronagraphs.
Classical optics (referred to as `bulk optics') are typically many thousands of times larger than the wavelength of light they shape and require precise and stable optomechanical components to accurately modify these wavefronts.
Integrated (or `photonic') optics enable direct manipulation of the complex electric fields at the scale of the wavelengths used.
Miniaturisation of previously discrete macro optics and their manufacture within a single homogeneous substrate removes both the requirement for separate optomechanical alignment and temperature related misalignment that is associated with their mechanical mounts.

Beam combiners that are required for optical and NIR interferometers require temperature and vibration controlled optical tables with sub-wavelength stability tolerances and alignment for the beamsplitters and associated optics.
Manufacture of waveguides within optical materials that perform the beam division and combination considerably simplify the optomechanical requirements, but then the challenges are in coupling the light from the macro optics into the substrates whilst keeping the transmitted efficiency high: diffraction limited optics are required to form \acp{psf} that couple efficiently into the near-single mode sized micro-optics.
Early examples include beam combiners for optical astronomical interferometers \citep[for example the IOTA/IONIC beam combiner; ][]{Berger01} and photonic lanterns, see \citet{Leon-Saval10} and references therein.
Typical coupling efficiencies are on the order of 10\%, increasing to 90\% for more recent designs, for example the efficient injection from large telescopes into single-mode fibres \citep{Jovanovic17}.
Full electromagnetic propagation is required to design and evaluate these photonic systems, but their complexity also enables new optical designs which can be combined to form compact, robust instrumentation, see the reviews in \citet{Minardi21,Jovanovic23}.

Coupling multi-mode light into monomode photonics is done using photonic lanterns, where a multimode input converted into the areal equivalent of a number of single mode optical channels, \citep{Norris22}.
A device equivalent to a Fabry-Perot etalon can be constructed by etching an elongated loop with one half of the loop parallel to the waveguide - frustrated transmission between the waveguide and the closed loop is modulated as a function of the number of integer wavelengths around the closed loop, imparting a precise frequency comb into the light.
A small heating element on top of the closed loop can change the length of the loop.
Due to the small physical size, this modulation can be in the kHz range.
All these photonic concepts are being considered for the design of coronagraphs for next generation space telescopes in order to image and characterise exoplanets, exploring concepts of different combinations of photonic and bulk optics \citep{Desai23a}.

\section{Algorithms for estimating the instantaneous PSF}

Deviations from the ideal optical prescription of telescope and instrument optics result in wavefront errors which manifest themselves as intensity deviations from the theoretical \ac{psf}.
Furthermore, these deviations change in intensity and position with time in the \ac{scfp}, and these can be equal to or larger than the flux from the astrophysical object next to the star.
The question is then how to estimate the science camera \ac{psf} for every single science camera exposure and subtract this estimate from the science camera image leaving only the flux from astrophysical objects adjacent to the target star.
This becomes more complicated when the position and brightness of the exoplanet is not known.
Several diversities - properties of the exoplanet that are not the same as the stellar halo - can differentiate between them.
The most important of these are:

\begin{itemize}
    \item Angular diversity: For an alt/az telescope, the planet on the sky has a predictable angular position and velocity with respect to the orientation of the instrument optics.
    \item Spectral diversity: The planet has a different spectral energy distribution, meaning that the relative flux between star and planet changes with wavelength.
    \item Polarimetric diversity: The light from star is almost completely unpolarized, but reflected light from clouds or dust around the exoplanet become polarsied under single scattering \citep{Gledhill91}.
    \item Wavelength diversity: The stellar halo scales with \ld{}, but the planet remains at the same location on the sky.
    \item Coherence diversity: The exoplanet flux is not coherent with the stellar halo and so does not interfere with it.
    \item Stochastic Speckle Discrimination: and the intensity fluctuations of the Airy core on ground based telescopes has a different statistical distribution \citep{Gladysz09}.
\end{itemize}

Many algorithms have been developed to take one or more of these diversities and provide estimates of the science camera \ac{psf}, using different linear combinations of the science camera images to estimate the instantaneous science camera \ac{psf}.
{\it Hubble Space Telescope} (HST) images of circumstellar material showed residual speckles that obscure the faint circumstellar environment, even after the subtraction of an image of a nearby star used as a reference \ac{psf}.
The concept of ``roll subtraction''  \citep{Schneider98} was used to estimate and remove these residual speckles.
Two or more images of the science target were taken with the telescope set at different angles about the target axis, so that the astronomical field would be rotated with respect to the (almost static) speckle field.
This was demonstrated in \citet{Schneider99} with the image of the disk around HR~4796A.
Even with the HST, the roll observations were taken within 25 minutes of each other to minimise changes in the telescope's optical path resulting from the ``breathing'' of the telescope optical assembly as it passed from day to night in its low earth orbit \citep{Bely93}.

With a ground based telescope, the speckle field changes on shorter timescales and with increased complexity because of (i) a continuously changing gravity vector on the telescope and instrument (ii) temperature and mechanical variations in the optomechanics within the instrument and (iii) changes in the performance of the adaptive optics system due to changing atmospheric conditions.
Angular Differential Imaging \citep[ADI; ][]{Marois06} has become a fundamental algorithm for many ground based telescope observations where significant sky rotation occurs during the observations of the planet.

Exploiting narrow band absorption features in the gas giant exoplanet spectrum enabled Methane Spectral Imaging (MSI), with TRIDENT \citep{Marois05} being one of the first cameras built to exploit this, along with the SDI camera at the MMT and VLT \citep{Biller07-1}.
With AO systems reaching to optical wavelengths, this has had a renaissance with H$\alpha$ imaging with the MUSE integral field spectrograph and the discovery of the accreting protoplanet PDS~70c \citep{Haffert19}.

Estimating the stellar halo with images at nearby wavelengths was generalised with the use of integral field spectrographs, where many science camera \acp{psf} are sampled at different wavelengths simultaneously to form $(x,y,\lambda)$ data cubes.
Resampling the image slices into the same \ld{} spatial scale radially smears out any exoplanet signal, so subtracting off a median of these images removes the stellar halo but keeps most of the planet flux intact, making it visible when the median subtracted cube is resampled into the sky coordinates and combined to produce the \ac{psf} subtracted image, generalised as Spectral Differential Imaging \cite[SDI; ][]{Sparks02} and demonstrated on-sky with SINFONI \citep{Thatte07}.
For higher spectral resolutions, the spectrum of the exoplanet begins to resolve the individual rotational-vibrational transitions, enabling High Spectral Contrast Imaging \citep[including the detection of HD~209458b ][]{Snellen10} and then generalising into the principle of molecule mapping in directly imaged exoplanets such as Beta~Pictoris~b \citep{Hoeijmakers18}.

Stochastic Speckle Discriminaton \citep[SSD; ][]{Gladysz09} is possible using short exposure images with the Airy core unsaturated.
Photon counting devices enable this detection method to work - this was demonstrated using an MKID detector and has led to the discovery of a substellar companion using this technique \citep{Steiger21}.
Each diversity improves sensitivity from a factor of a few to ten or more.
Combining different diversities together results in an even greater cumulative effect.

\section{Conclusions}

Since the first detection of planets outside our solar system with the pulsar planets \citep{Wolszczan92} using pulsar timing, and the first exoplanet around a solar-type star \citep[51 Peg b; ][]{Mayor95} using radial velocity measurements on the star, we now have thousands of planets indirectly detected with radial velocity and transit methods.
Direct imaging of exoplanets have revealed dozens of young, self-luminous gas giant planets \citep{Currie23,Chauvin24} and with the minimized infrared background accessible with the \ac{jwst}, we are entering the era of directly detecting sub-Jupiter mass planets.

\acp{elt} with extreme AO systems and the next generation of space telescopes enable the reflected light detection of planets around the nearest stars.
The direct imaging of exoplanets is a dynamic and rapidly changing field, with each decade of suppression bringing new challenges and researchers searching for (and finding) solutions to them.
We have developed the mathematical theory that describes coronagraphs, verified them in the laboratory, and demonstrated them on sky; we also develop algorithms to tease out these faint signals against the almost overwhelming glare of their parent stars.
It is perhaps inevitable that in the next decade we will be imaging and characterising pale blue dots around our nearest neighbours, and we will take one further step towards seeing if the Earth is truly unique.

\section*{DISCLOSURE STATEMENT}
The authors are not aware of any affiliations, memberships, funding, or financial holdings that might be perceived as affecting the objectivity of this review.

\section*{ACKNOWLEDGMENTS}
This work benefited from the 2023 Exoplanet Summer Program in the Other Worlds Laboratory (OWL) at the University of California, Santa Cruz, a program funded by the Heising-Simons Foundation and NASA.

To achieve the scientific results presented in this article we made use of the \emph{Python} programming language\footnote{Python Software Foundation, \url{https://www.python.org/}}, especially the \emph{SciPy} \citep{virtanen2020}, \emph{NumPy} \citep{numpy}, \emph{Matplotlib} \citep{Matplotlib}, \emph{emcee} \citep{foreman-mackey2013}, and \emph{astropy} \citep{astropy_1,astropy_2} packages.
The \ac{hci} simulations were calculated with \emph{HCIpy} \citep{Por18}.
This Article is reproducible using the \emph{showyourwork!} workflow management tool \citep{Luger2021}.
This research has made use of NASA's Astrophysics Data System Bibliographic Services.
M.\ A.\ K.\ acknowledges useful conversations with several people (most notably
Phil Hinz, Eric Mamajek and Andrew Skemer) at the Humble Sea Brewing Company.

\bibliographystyle{ar-style2}
\bibliography{bib}

\section*{RELATED RESOURCES}

This manuscript was prepared using the \project{showyourwork!} package\footnote{\url{https://show-your.work}} and the source code used to generate each figure is available in a public \project{GitHub} repository\footnote{\url{https://github.com/mkenworthy/ARAA_HCC}}.

\end{document}